\documentclass[12pt]{article}
\pdfoutput=1

\usepackage[utf8]{inputenc}

\usepackage{graphicx,psfrag,epsf,color}
\usepackage{amsmath,amssymb,amsfonts}
\usepackage{array}
\usepackage{cite}
\usepackage{rotating}
\usepackage{slashed, cancel}
\usepackage{scalerel,stackengine}
\usepackage{multirow}

\numberwithin{equation}{section}

\newcommand{\be}{\begin{equation}}
\newcommand{\ee}{\end{equation}}
\newcommand{\bea}{\begin{eqnarray}}
\newcommand{\eea}{\end{eqnarray}}
\newcommand{\bi}{\begin{itemize}}
\newcommand{\ei}{\end{itemize}}
\newcommand{\ben}{\begin{enumerate}}
\newcommand{\een}{\end{enumerate}}
\newcommand{\bt}{\begin{tabular}}
\newcommand{\et}{\end{tabular}}

\newcommand{\nn}{\nonumber}

\newcommand{\nm}{n_-}
\newcommand{\np}{n_+}

\newcommand{\T}{{\bf T}}

\newcommand{\ep}{\epsilon}
\newcommand{\bz}{\bar{z}}

\newcommand{\as}{\alpha_s}
\newcommand{\eps}{\epsilon}
\newcommand{\ord}{{\cal O}}

\newcounter{MBQ}

\newcounter{MBA}

\begin{document}
\allowdisplaybreaks

\begin{titlepage}

\begin{flushright}
{\small
TUM-HEP-1270/20\\
CERN-TH-2020-135\\
11 August 2020
}
\end{flushright}

\vskip0.6cm
\begin{center}
{\Large \bf\boldmath 
Large-$x$ resummation of off-diagonal deep-inelastic\\[0.0cm]  
parton scattering from $d$-dimensional refactorization\\[0.25cm] 
}
\end{center}

\vspace{0.2cm}
\begin{center}
{\sc Martin~Beneke},$^{a}$ 
{\sc Mathias~Garny},$^{a}$
{\sc Sebastian Jaskiewicz},$^{a}$ \\ 
{\sc Robert~Szafron},$^{b}$
{\sc Leonardo Vernazza},$^{c}$ 
and {\sc Jian~Wang}$^{d}$\\[6mm]
{\it $^a$Physik Department T31,\\
James-Franck-Stra\ss e~1, 
Technische Universit\"at M\"unchen,\\
D--85748 Garching, Germany\\[0.2cm]
$^b$Theoretical Physics Department, CERN,\\ 
1211 Geneva 23, Switzerland\\[0.2cm]
$^c$Dipartimento di Fisica Teorica, 
Universit\`a di Torino, \\
and INFN, Sezione di Torino, Via P. Giuria 1, 
I-10125 Torino, Italy \\[0.2cm]
$^d$School of Physics, Shandong University,\\ 
Jinan, Shandong 250100, China\\[0.2cm]
}
\end{center}

\vspace{0.2cm}
\begin{abstract}
\vskip0.2cm\noindent
The off-diagonal parton-scattering channels $g+\gamma^*$ and
$q+\phi^*$ in deep-inelastic scattering are power-suppressed 
near threshold $x\to 1$. We address the next-to-leading power 
(NLP) resummation of large double logarithms of $1-x$ to all orders 
in the strong coupling, which are present even in the off-diagonal 
DGLAP splitting kernels.  The appearance of divergent convolutions 
prevents the application of factorization methods
known from leading power resummation.
Employing $d$-dimensional consistency relations 
from requiring $1/\epsilon$ pole cancellations in 
dimensional regularization between momentum regions, 
we show that the resummation of the 
off-diagonal parton-scattering channels at the leading logarithmic 
order can be bootstrapped from the recently conjectured 
exponentiation of NLP  soft-quark Sudakov 
logarithms. In particular, we derive 
a result for the DGLAP kernel in terms of the series of 
Bernoulli numbers found previously by Vogt directly from algebraic 
all-order expressions. We identify the off-diagonal DGLAP 
splitting functions and soft-quark Sudakov logarithms as 
inherent two-scale quantities in the large-$x$ limit. 
We use a refactorization of these scales and renormalization 
group methods inspired by soft-collinear effective 
theory to derive the  conjectured soft-quark Sudakov
exponentiation formula. 
\end{abstract}
\end{titlepage}

\section{Introduction}
\label{sec:introduction}

Resummations of logarithmically enhanced loop corrections 
are a powerful and often essential 
tool to enlarge the predictivity of QCD perturbation theory.
Resummation is necessary when a ratio of kinematic 
invariants, $\lambda$, becomes small such that 
$\alpha_s\ln^k\lambda$, where $\alpha_s$ is the strong 
coupling and $k=1$ or $2$, is no longer a good expansion 
parameter. Recent interest in this subject has focused 
on understanding the structure of such logarithmic terms 
at next-to-leading power (NLP) in $\lambda$ with the aim 
of summing them to all orders in $\alpha_s$. This has 
been accomplished at the leading-logarithmic (LL) order in 
various contexts, covering final-state event shapes 
\cite{Moult:2018jjd,Moult:2019vou}, threshold resummation 
in Drell-Yan and Higgs production 
\cite{Beneke:2018gvs,Beneke:2019mua,Bahjat-Abbas:2019fqa}, 
and Higgs production or decay through light-quark loops 
\cite{Liu:2019oav,Wang:2019mym,Anastasiou:2020vkr}. 
A number of methods has been used, but it has become 
evident that a generalization to the next-to-leading-logarithmic 
(NLL) order is not straightforward. This is to be compared to 
the situation at leading power (LP), where resummation 
is often understood to any logarithmic order, even though one 
faces technical challenges of high-order loop calculations 
in practice.

The most natural framework to formulate resummation is 
through the factorization of scales and evolution equations. 
The all-order resummed expression is then obtained as 
the product or convolution of the factorized pieces. 
At NLP, one faces the new difficulty that these convolutions 
are divergent. While divergent convolutions are familiar 
from rapidity divergences, which are not regulated dimensionally 
and may occur already at LP, such as in transverse-momentum 
dependent factorization, the problem at NLP is of a 
different nature. The divergences 
can be regulated dimensionally and arise in convolutions of 
factors containing the physics at different virtualities.  
However, since factorization and resummation refer to the 
renormalized factors before convolution, the standard 
formalism fails to deal with this situation.
An explicit example can be found in \cite{Beneke:2019oqx} 
for the case of next-to-leading logarithms near the 
$q\bar{q}\to \gamma^*$ Drell-Yan threshold.

In the present paper, we address these difficulties for 
the threshold of off-diagonal deep-inelastic parton
scattering. The off-diagonal channels vanish at LP near 
threshold $x\to 1$, since they do not contain $1/[1-x]_+$ 
distributions at any order in $\alpha_s$. However, the failure 
of standard resummation methods appears already at the 
LL order for the DGLAP splitting functions. Vogt and 
collaborators \cite{Vogt:2010cv,Almasy:2010wn,Vogt:2012gb}
found that the all-order quark-gluon splitting function 
with LL accuracy is given in moment space by 
\begin{equation}
P_{gq}^{\rm LL}(N) = \frac{1}{N}\frac{\alpha_s C_F}{\pi}\,
\mathcal{B}_0(a), 
\qquad  a = \frac{\alpha_s}{\pi} (C_F-C_A) 
\ln^2 N\,,
\label{eq:vogtconjecture}
\end{equation}
where
\begin{equation}
\mathcal{B}_0(x) = \sum_{n=0}^\infty\frac{B_n}{(n!)^2}x^n
\label{eq:B0fn}
\end{equation}
is the Borel transform of the generating function of the 
Bernoulli numbers $B_0 = 1$, $B_1=-1/2$, $\dots$.\footnote{Here 
and in the remainder of the paper $\alpha_s$ without argument denotes 
the strong coupling in the $\overline{\rm MS}$ scheme 
at the renormalization / factorization scale $\mu$. 
With our definition (\ref{eq:gamgqabstract}) of the anomalous 
dimensions, the splitting kernel differs by a factor of two 
from \cite{Vogt:2010cv}.} 
The existence of an infinite series of double logarithmic 
terms shows that in the off-diagonal channels even the 
anomalous dimension is a two-scale quantity as $N\gg 1$, 
contrary to the diagonal anomalous dimension, a distinction 
that has not received as much attention as it deserves, 
except for \cite{Vogt:2010cv,Almasy:2010wn,Vogt:2012gb}. 
This remarkable result was obtained from the structure of 
$1/\epsilon$ poles of the unfactorized parton-scattering 
cross sections in the exactly known $d=4-2\epsilon$ dimensional 
low-order results, and 
their consistency with factorization. These structures were 
then extrapolated to all orders to find closed functional 
forms, including the reconstruction of the series of 
Bernoulli numbers. To our understanding (\ref{eq:vogtconjecture}) 
has not yet been proven by deriving it directly from 
algebraic all-order expressions.
With this method, further results on the 
finite short-distance coefficients of the off-diagonal 
channels in deep-inelastic scattering 
and Drell-Yan production at large $N$ were also obtained  
\cite{Vogt:2010cv,Almasy:2010wn}.

What distinguishes the off-diagonal splitting functions from 
the diagonal ones in the $x\to 1$ / large-$N$ limit is that 
the former describe the splitting into an energetic parton 
and a soft {\em quark}. We further notice that the 
double-logarithmic series involves the colour factor 
$C_F-C_A$. A connection between soft quarks and this 
colour factor of the large logarithms also appears for the 
Sudakov resummation of the  $q\bar{q} \to \phi^*$ form factor, 
where $\phi^*$ denotes a Higgs boson, effectively coupled to 
two gluons, and $q$ a light quark~\cite{Liu:2017vkm,Liu:2018czl}. 
The leading logarithms here 
originate from soft quark exchange. Further, 
the authors of \cite{Moult:2019uhz} investigated the 
all-order structure of the  $e^+ e^- \to q\bar{q}g$ amplitude 
in the kinematic configuration where the quark and anti-quark 
are nearly collinear with small virtuality $s\ll Q^2$ and 
momentum fractions $z$ and $\bar z=1-z$, respectively, 
recoiling against the energetic gluon. Keeping the leading 
double poles $1/\epsilon^2$ and logarithms of $z$ or $\bar z$, 
as the quark or anti-quark become soft, they conjectured 
the exponentiation of the corresponding one-loop terms 
to all orders 
\bea
\mathcal{P}_{q\bar q}(z) &=& 
\mathcal{P}^{\rm tree}_{q\bar q}(z) \,
\exp\Bigg[\frac{\alpha_s}{\pi\epsilon^2}\,
\Bigg\{\T_1\cdot\T_3\left(\frac{\mu^2}{zQ^2}\right)^{\!\epsilon}
+ \T_2\cdot\T_3\left(\frac{\mu^2}{\bar{z}Q^2}\right)^{\!\epsilon}
+ \T_1\cdot\T_2\left(\frac{\mu^2}{s}\right)^{\!\epsilon}
\nn\\
&+& 
\T_1\cdot\T_2\left(\left(\frac{\mu^2}{Q^2}\right)^{\!\epsilon}
- \left(\frac{\mu^2}{z\bar{z}Q^2}\right)^{\!\epsilon}\,\right)
-
\T_1\cdot\T_2\left(\left(\frac{\mu^2}{s}\right)^{\!\epsilon}
- \left(\frac{\mu^2}{z\bar{z}s}\right)^{\!\epsilon}\,\right)
\Bigg\}\Bigg]\,.\quad
\label{eq:quarksudakovconjecture}
\eea
Here $\T_1$, $\T_2$, $\T_3$ are the colour operators of the quark, 
antiquark and gluon, respectively, such that 
$\T_1\cdot\T_2=C_A/2-C_F$, $\T_1\cdot\T_3=\T_2\cdot\T_3=-C_A/2$.
If we now take $z\to 0$, which corresponds to the quark 
becoming soft, and focus on the terms involving $Q^2$, 
we see that the coefficient of $Q^{-2\epsilon}$ has the 
colour factor $\T_2\cdot\T_3+\T_1\cdot\T_2=-C_F$.  The coefficient 
of $(zQ^2)^{-\epsilon}$, which involves the new scale 
$\sqrt{z}Q \ll Q$ in the soft-quark limit, however, is   
$\T_1\cdot\T_3-\T_1\cdot\T_2=C_F-C_A$. It is tempting to 
conjecture that in a splitting $1\to 2+3$ with soft 
3, the endpoint divergence, which occurs when integrating 
over $z$ since $\mathcal{P}^{\rm tree}_{q\bar q}(z) \propto 1/z$, 
and which requires extra resummation 
of the logarithms of $z$ or $\bar z$ not captured by the usual 
formalism, is related to the difference of the 
Casimir charge of the energetic particles 1 and 2. 
Along this line, it was noted in \cite{Moult:2019uhz,Moult:2019vou} 
that in supersymmetric QCD with quarks in the adjoint 
representation, the endpoint divergences and extra logarithms 
are absent. All three examples of the appearance of 
$C_F-C_A$ in front of double logarithms have in common that 
the resummed result was obtained without explicit 
factorization of the scales involved, using either $d$-dimensional 
arguments, diagrammatic arguments, or a conjecture.

In this paper, we establish a connection between some of these 
results in the context of NLP LL resummation for off-diagonal 
deep-inelastic parton scattering as $x\to 1$. To this 
end we adapt the soft-quark Sudakov exponentiation 
conjecture \cite{Moult:2019uhz} from event shapes 
to deep-inelastic 
scattering (DIS). We then
\begin{itemize}
\item prove (\ref{eq:vogtconjecture}) for the resummed 
off-diagonal splitting function and the finite coefficient 
function from the soft-quark Sudakov exponentiation 
conjecture via $d$-dimensional consistency relations 
that follow from the requirement of pole cancellation 
between momentum regions. The adapted version of 
(\ref{eq:quarksudakovconjecture}) plays the role of 
a ``boundary condition'' in the purely hard contribution 
to the process, from which the resummation of the 
full process follows in closed all-order form.
\item derive the previously conjectured 
exponentiation formula through the refactorization of 
certain power-suppressed operators in soft-collinear 
effective theory (SCET) which have endpoint-singular matching 
coefficients. The renormalization group equations (RGEs) 
then exhibit the origin of the  $C_F-C_A$ colour factor. 
\end{itemize}

The outline of the paper is as follows. 
In Section~\ref{sec:quarksudakov} we consider the 
$q+\phi^*\to q+g$ amplitude, define the light-cone momentum 
distribution for the $qg$-final state, and 
calculate its leading poles at the one-loop order. 
We then apply the exponentiation conjecture analogous to 
(\ref{eq:quarksudakovconjecture}) to the soft-quark limit 
of this amplitude. Sections~\ref{sec:consistencyrel} and 
\ref{sec:derivation} contain the material related to the two 
bullet points above, respectively. We conclude in 
Section~\ref{sec:conclusion}. Appendices~\ref{app:SCET} 
and \ref{app:largexDIS} collect SCET conventions and the 
field modes which appear in different parts of the paper, 
and some basic facts on DIS at large $x$. 
In Appendices~\ref{app:alt3.12} and~\ref{app:altB1} we provide 
alternative derivations of a) the solution of the consistency 
relations at LP, and b) the resummation of the refactorized SCET 
operator, which confirms the result of 
Section~\ref{sec:derivation}. The application of consistency 
relations to the thrust event shapes  considered in 
\cite{Moult:2019uhz} is presented in Appendix~\ref{app:eventshapes} 
in order to note the similarities and differences 
between the two processes.

\section{Off-diagonal DIS cross section and
soft-quark Sudakov exponentiation}
\label{sec:quarksudakov}

We begin by considering deep-inelastic scattering
\begin{equation}
q(p)+ \phi^*(q) \to X(p_X)
\end{equation}
of a quark off a Higgs boson, which couples to
quarks and gluons through the gluonic interaction
\begin{equation}
{\cal L} = \kappa\,  \phi\,
{\rm tr}\left[G^{\mu\nu}G_{\mu\nu}\right]\,.
\label{eq:Higgsvertex}
\end{equation}
The coupling $\kappa$ is related to the
effective Higgs-gluon coupling in the infinite
top-quark mass limit.\footnote{See \eqref{kappaDef} in
Appendix \ref{app:largexDIS}. We refer to this
appendix for a summary of
the kinematics of DIS, the factorization of the hadronic
structure function at leading power, and the relevant 
momentum regions at large $x$.}
We are interested in the kinematic situation
when the final state has small invariant mass,
$p_X^2\ll Q^2\equiv -q^2$, which
corresponds to the limit $x\to 1$ for the Bjorken scaling variable
$x\equiv Q^2/(2 p\cdot q)$.
The reason for considering
this exotic process is that it is related to Drell-Yan
production of a Higgs boson in the quark-gluon channel near
the partonic threshold. The above DIS process allows us
to extract the quark-gluon splitting kernel $P_{gq}$
\cite{Vogt:2010cv}, which enters Higgs production. All these
quantities are NLP, i.e. suppressed by one power of $(1-x)$
as $x\to 1$ relative to the leading diagonal gluon-gluon coupling.
The off-diagonal channel $g+\gamma^* \to q+g$ for the
more standard scattering on the vector current
can be obtained by substitution of colour factors
\cite{Vogt:2010cv}.

We consider the dimensionally regularized and
unfactorized partonic DIS cross sections. 
Following \cite{Vogt:2010cv}, we introduce the partonic 
structure functions $W_{\phi,g}$, $W_{\phi,q}$
by defining 
\bea \label{hadrtensdefPartonic}
W_{\phi,i} = \frac{1}{8\pi Q^2} \int d^4 x \, e^{i q \cdot x} \,
\big\langle i(p) \big| \big[G_{\mu\nu}^A G^{\mu\nu A}\big](x)
\big[G_{\rho\sigma}^B G^{\rho\sigma B} \big](0) \big| i(p) 
\big\rangle,
\eea
where $i = g$ or $q$. 
At the lowest non-vanishing order in $\alpha_s$, to NLP
in $(1-x)$, and neglecting ${\cal O}(\epsilon^0)$ terms not
multiplied by logarithms of $(1-x)$ one has
\begin{equation}
W_{\phi,g}=\delta(1-x)+\mathcal{O}(\alpha_s), \qquad
W_{\phi,q}=-\frac{1}{\epsilon}
\, \frac{\alpha_s C_F}{2\pi}\,(1-x)^{-\epsilon} +
\mathcal{O}(\alpha_s^2) \,.
\label{eq:LOTs}
\end{equation}
The exact result is provided in \eqref{WphiTreeB} 
and \eqref{Wsq1Full}. 

\subsection{Momentum distribution
function and lowest-order result}

\begin{figure}[t]
\begin{center}
  \includegraphics[width=0.28\textwidth]{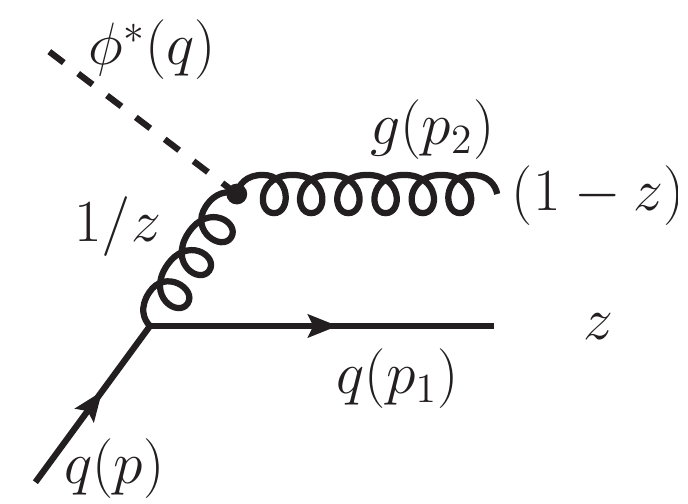}
\end{center}
\caption{Scattering of a quark off a virtual Higgs boson at
tree level.}
\label{fig:treescattering}
\end{figure}

At the lowest order the quark-scattering off the Higgs boson is
realized by the process
\begin{equation}
q(p) + \phi^*(q) \to q(p_1)+g(p_2)\,.
\label{eq:DISvirtual}
\end{equation}
The tree-level amplitude is shown in
Figure~\ref{fig:treescattering}. We write the
$d=4-2\epsilon$ dimensional two-particle phase
space as
\bea\label{PS2def}
d\Phi_{2}(p_X;p_1,p_2)&=&
\frac{\tilde\mu^{2\epsilon}d^dp_1}{(2\pi)^d}\, 2\pi \delta(p_1^2)
\theta(p_1^0)\, 2\pi\delta\big[(p_X-p_1)^2\big]\theta(p_X^0-p_1^0)
\nn\\
&=& \frac{e^{\gamma_E\epsilon}}{8\pi\Gamma(1-\epsilon)}\,d z
\left(\frac{\mu^2}{s_{qg}z\bar z}\right)^{\!\epsilon}
\theta(z)\theta(\bar z)  \theta(s_{q g})\,,
\eea
where $\tilde\mu^2 = \mu^2 e^{\gamma_E}/(4 \pi)$,
$s_{qg}=p_X^2$. We introduced the variable
\be\label{zDefB}
z \equiv \frac{n_- p_1}{n_- p_1 + n_- p_2}\, ,
\qquad \qquad \quad
\bar z=1-z\, ,
\ee
which represents the distribution of the light-cone final-state
momentum between the nearly collinear quark
and gluon in the final state. The vector $n_-$
represents a light-like vector which projects on the large momentum 
components of the final state particles.\footnote{See  Appendix 
\ref{app:SCET} for more details on the definition 
of SCET reference vectors and of the power counting parameters
relevant for DIS near threshold.}

We can represent the matrix element squared of the process 
(\ref{eq:DISvirtual}) integrated over the phase space in the form
\be
\label{eq:TphiqVsz}
W_{\phi,q}\big|_{q\phi^*\to qg} = \int_0^1 dz\,
\left(\frac{\mu^2}{s_{qg}z\bar z}\right)^{\!\epsilon}
\mathcal{P}_{qg}(s_{qg},z)
\Big|_{s_{qg}=Q^2\frac{1-x}{x}}\, ,
\ee
which holds to any order in perturbation theory in the strong
coupling $\alpha_s$. The momentum distribution function
is defined as
\be
\label{eq:defPqg}
\mathcal{P}_{qg}(s_{qg},z) \equiv
\frac{e^{\gamma_E\epsilon}\, Q^2}{16\pi^2\Gamma(1-\epsilon)}
\frac{|{\cal  M}_{q\phi^*\to qg}|^2}{|{\cal M}_0|^2}\,,
\ee
where $|{\cal M}_0|^2$ denotes the
tree-level matrix element squared, averaged (summed) over
the spin and colour of the initial (final) state 
for the leading  diagonal 
channel process $g+\phi^*\to g$. For $x \to 1$, we can 
expand $\mathcal{P}_{qg}(s_{qg},z)$ in $s_{qg}/Q^2$ or 
$\lambda \sim \sqrt{1-x} \ll 1$. For this purpose we note that
\be
\label{eq:sqg}
s_{qg} = (p+q)^2 = \frac{Q^2(1-x)}{x} = Q^2(1-x) 
+ \ord(\lambda^4)\,.
\ee
From \eqref{MggTree} we have
\be
|{\cal M}_0|^2 \equiv \big| {\cal M}_{g \phi^{*}  \to g} \big|_{\rm tree}^{2}
 = \frac{\kappa^2  Q^4}{x^2}\, (1-\eps)
 = \kappa^2  Q^4\, (1-\eps) + \ord(\lambda^2)\,.
\ee
Similarly, the expansion of $|{\cal  M}_{q\phi^*\to qg}|^2$, 
given in \eqref{Mqg1}, which is itself a function of 
$s_{qg}$ or $x$ and of the momentum fraction $z$, 
gives 
\be\label{Mqg1expanded}
\big| {\cal M}^{(1)}_{q \phi^{*}  \to qg} \big|^{2} =
2 \kappa^2 \, g_s^2 C_F\, (1-\eps) \, Q^2 \frac{\bar z^2}{z} + \ord(\lambda^2)
\ee
at the lowest non-vanishing order in the coupling expansion,  
which implies
\be\label{eq:Pqgtree}
\mathcal{P}_{qg}(s_{qg},z)\big|_{\rm tree} =
\frac{\alpha_sC_F}{2\pi}\frac{\bar z^2}{z}  + 
\mathcal{O}(\epsilon, \lambda^2) \, .
\ee
Integrating and neglecting ${\cal O}(\epsilon)$
corrections that are \emph{not} multiplied by
logarithms (i.e. counting $\epsilon\ll 1$ but
$\epsilon\ln(1-x)\sim 1$, $(1-x)^{-\epsilon}\sim 1$ and 
$\epsilon\ln(\mu/Q) \sim 1$), gives
\be
 W_{\phi,q}\big|^{\rm NLP}_{\ord(\as), \, \rm leading\,pole}
 = -\frac{1}{\epsilon} \,\frac{\alpha_sC_F}{2\pi}
 \left(\frac{\mu^2}{Q^2(1-x)}\right)^{\!\epsilon},
\ee
in agreement with (\ref{eq:LOTs}). 
$W_{\phi,q}\big|_{q\phi^*\to qg}$ represents the contribution
to the {\em partonic} DIS structure function when only two
partons are present in the final state. As such it is an
infrared (IR) divergent quantity. In lowest order in $\alpha_s$,
the IR divergence is a single $1/\epsilon$ pole,
which arises from the $z\to 0$ region of the integral
(\ref{eq:TphiqVsz}) owing to the $1/z$ behaviour of the tree-level 
momentum distribution function. 
The $z\to 0$ limit corresponds to the kinematic
configuration where the initial quark transfers all of its
momentum to the final-state gluon, and the final-state quark
becomes soft. It is therefore essential that the integration over
$z$ in (\ref{eq:TphiqVsz}) is done in $d$ dimensions, a fact that
will be of importance later on.

\subsection{One-loop momentum distribution
function}

\begin{figure}[t]
\begin{center}
  \includegraphics[width=0.80\textwidth]{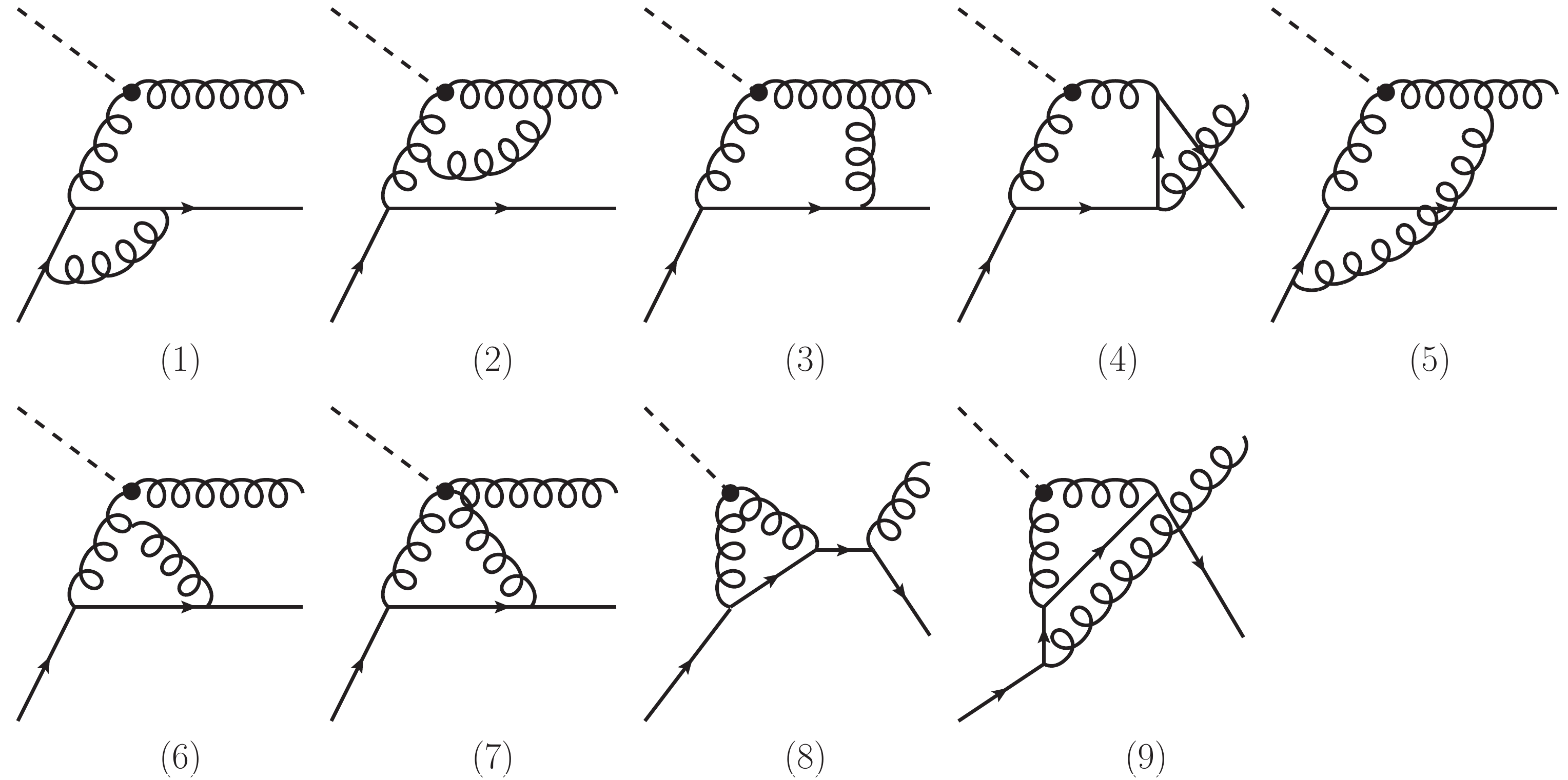}
\end{center}
\vspace*{-0.2cm}
\caption{One-loop corrections to the scattering of a quark 
off a virtual Higgs boson. Only the triangle and box diagrams 
are shown.}
\label{fig:oneloopP}
\end{figure}

In this subsection, we calculate the one-loop (virtual) 
correction to the $2\to 2$ scattering process 
$q\phi^*\to qg$ and obtain the corresponding 
momentum distribution function (\ref{eq:defPqg}).
We are interested in the leading double poles and logarithms.
We therefore need the leading pole $1/\epsilon^2$ without 
expanding its coefficient in a series of $\epsilon$.
The relevant Feynman diagrams are shown in Figure~\ref{fig:oneloopP}.
We find that only the first five diagrams give 
non-vanishing leading poles. 
Calculating the interference of these one-loop diagrams with 
the tree diagram, we obtain for the leading-pole terms
\begin{align}
\mathcal{P}_{qg}(s_{qg},t,u)|_{\rm 1-loop}& = 
\mathcal{P}_{qg}(s_{qg},t,u)|_{\rm tree}  \frac{\as}{\pi} 
\frac{\mu^{2\ep}}{\ep^2}\,\bigg\{
\nn\\
&\left[ \T_1 \cdot\T_2\,(-s_{qg})^{-\ep} +  
\T_1 \cdot\T_0\,(-t)^{-\ep}  +  
\T_2 \cdot\T_0\,(-u)^{-\ep} \right] \nn \\
& +  \T_1 \cdot\T_2 \left( (Q^2)^{-\ep}-(-t)^{-\ep} - (-s_{qg})^{-\ep}+\frac{(-s_{qg})^{-\ep}(-t)^{-\ep}}{(Q^2)^{-\ep}} \right)\nn\\
& +  \T_1 \cdot\T_0 \left( (Q^2)^{-\ep}-(-s_{qg})^{-\ep} - (-u)^{-\ep}+\frac{(-s_{qg})^{-\ep}(-u)^{-\ep}}{(Q^2)^{-\ep}} \right)\nn\\
& +  \T_2 \cdot\T_0 \left( (Q^2)^{-\ep}-(-t)^{-\ep} - (-u)^{-\ep}+\frac{(-t)^{-\ep}(-u)^{-\ep}}{(Q^2)^{-\ep}} \right)\bigg\}
\label{eq:P1loopstu}
\end{align}
with $t=(p_1-p)^2$, $u=(p_2-p)^2$, and
\begin{align}
\mathcal{P}_{qg}(s_{qg},t,u)|_{\rm tree}= 
\frac{\alpha_sC_F}{2\pi}\frac{s_{qg}^2+u^2}{-tQ^2} + \ord(\epsilon)\,,
\end{align}
employing colour operator notation, $\T_i$, for particle $i$ 
($i=0$ for the incoming quark). 
Eq.~(\ref{eq:P1loopstu}) is valid for general values of the 
Mandelstam variables $s_{qg},t,u$.
The first line in the curly bracket is the familiar double-pole 
structure when all $s_{qg},t,u$ are of order of the hard scale $Q^2$.
The last three lines represent additional contributions, which 
are suppressed by $\epsilon^2$ compared to the first line. 
Although technically finite as $\epsilon\to 0$, 
these cannot be simply omitted because after integration
over the phase space to obtain $W_{\phi,q}$, 
they may generate poles and logarithms of equal order as the 
pole terms from the first line, as will be shown below.

When the final-state quark and gluon become collinear, 
$s_{qg}\ll Q$. With $s_{qg}+t+u=-Q^2$, we can 
parameterize $t=-zQ^2-s_{qg}/2, u=-\bar{z}Q^2-s_{qg}/2$. 
The collinear expansion amounts to expanding 
in $s_{gq}/Q^2 \sim  (1-x)\sim \lambda^2$ at fixed $z$, 
which yields 
\begin{align}
\mathcal{P}_{qg}(s_{qg},z)|_{\rm 1-loop}& = 
\mathcal{P}_{qg}(s_{qg},z)|_{\rm tree}\frac{\as}{\pi} 
\frac{\mu^{2\ep}}{\ep^2}\,\bigg\{
\nn\\
& \left[ \T_1 \cdot\T_2\,(-s_{qg})^{-\ep} 
+  \T_1 \cdot\T_0\,(zQ^2)^{-\ep}  +  
\T_2 \cdot\T_0\,(\bz Q^2)^{-\ep} \right] \nn \\
& +  \T_1 \cdot\T_2 \left[ (Q^2)^{-\ep}-(zQ^2)^{-\ep} - (-s_{qg})^{-\ep}+(-zs_{qg})^{-\ep} \right]\nn\\
& +  \T_1 \cdot\T_0 \left[ (Q^2)^{-\ep}- (\bz Q^2)^{-\ep}-(-s_{qg})^{-\ep} +(-\bz s_{qg})^{-\ep} \right]\nn\\
& +  \T_2 \cdot\T_0 \left[ (Q^2)^{-\ep}-(zQ^2)^{-\ep} - (\bz Q^2)^{-\ep}+(z \bz Q^2)^{-\ep} \right]\bigg\}\,.
\label{oneloop-coll}
\end{align}
Given that the leading order result \eqref{Mqg1expanded} becomes singular only at the end point $z=0$,
we can safely expand the $\bz^{-\epsilon}$ terms in a series of $\epsilon$ in the above equation.
Therefore, keeping only terms contributing to the leading poles after integration over the phase space, we have
\bea
\mathcal{P}_{qg}(s_{qg},z)|_{\rm 1-loop} &=&
\mathcal{P}_{qg}(s_{qg},z)|_{\rm tree}\,
\frac{\alpha_s}{\pi}\frac{1}{\epsilon^2}\,
\bigg(\T_1\cdot\T_0\left(\frac{\mu^2}{ zQ^2}\right)^{\!\epsilon}
  + \T_2\cdot\T_0\left(\frac{\mu^2}{\bz Q^2}\right)^{\!\epsilon}
\nn\\
&&
\hspace*{-2cm}
+ \,\T_1\cdot\T_2\left[\left(\frac{\mu^2}{Q^2}\right)^{\!\epsilon}
- \left(\frac{\mu^2}{z Q^2}\right)^{\!\epsilon}
+\left(\frac{\mu^2}{z s_{qg}}\right)^{\!\epsilon}\,\right]
\bigg)\,.
\label{eq:Pqg1loop}
\eea
In the framework of SCET  
this result exhibits a profound problem. The tree amplitude 
represented in Figure~\ref{fig:treescattering} corresponds 
to a  $J^{B1}$ SCET operator (see Appendix~\ref{app:SCET})
with a quark field in 
the collinear direction, and a quark and a gluon field 
in the anti-collinear direction with light-cone 
momentum fractions $z$ and $\bar{z}$, respectively. The tree-level 
matching coefficient of this operator from the diagram in 
Figure~\ref{fig:treescattering} is proportional to $1/z$, 
which gives the $1/z$ behaviour of 
$\mathcal{P}_{qg}(s_{qg},z)|_{\rm tree}$ after squaring 
the amplitude and accounting for a factor of $z$ from 
the sum of the final-state quark spin. From the 
general formula for the anomalous dimension of subleading 
power operators
\cite{Beneke:2017ztn,Beneke:2018rbh}, we get the double pole terms
with $\T_1\cdot\T_0$ and $\T_2\cdot\T_0$ from the standard cusp 
anomalous dimension terms. However, one cannot obtain a
cusp term for the two fields within the same collinear sector, 
i.e.~the $\T_1\cdot\T_2$ term.
In this part, there are three terms involving three 
different scales. The third contains the scale $zs_{qg}$. 
We may disregard it here because the dependence on 
$s_{qg}$ identifies it as a term related to the final-state 
jet function, rather than the renormalization of the 
$J^{B1}$ operator  at the hard DIS vertex.
The first two terms, however, contain the hard scales 
$Q^2$ and $zQ^2$, and they are supposed to be predicted by 
the corresponding anomalous dimension. However, the 
anomalous dimension given in \cite{Beneke:2017ztn,Beneke:2018rbh} 
applies when the convolution of the coefficient function 
with the anomalous dimension is convergent, which is not 
the case here as discussed next.
 
The difference between these two terms in the coefficient 
of $\T_1\cdot\T_2$ is $\mathcal{O}(\epsilon)$ and hence 
does not contribute to the double pole. Instead, the 
expansion in $\epsilon$ produces $1/\epsilon\times \ln z$. 
The important point is that the $1/z$ singularity 
of the matching coefficient promotes this term to the same 
leading-pole order $1/\epsilon^3$ as the standard 
double pole terms after integration over $z$ as in 
\eqref{eq:TphiqVsz}. Moreover, 
the integral over $z$ must itself be regularized 
due to the singularity at $z=0$, and the correct result 
is obtained by {\em not} expanding (\ref{eq:Pqg1loop}) 
before integration. This can easily be seen by comparing 
(no expansion before integration)
\begin{equation}
\frac{1}{\epsilon^2} 
\int_0^1dz\,\frac{1}{z^{1+\epsilon}}\,(1-z^{-\epsilon})
=-\frac{1}{2\epsilon^3}
\end{equation}
to (expansion of (\ref{eq:Pqg1loop}) before integration)
\begin{equation}
\frac{1 }{\epsilon^2} \int_0^1dz\,\frac{1}{z^{1+\epsilon}}\, 
\left(\epsilon\ln z -\frac{\epsilon^2}{2!}\ln^2 z+\frac{\epsilon^2}{3!}\ln^3 z+\cdots  \right)
=-\frac{1}{\epsilon^3}+\frac{1}{\epsilon^3}-\frac{1}{\epsilon^3}
+\cdots\,.
\end{equation}
If only the pole part of the integrand were kept, the result would 
be incomplete. This explains why it was necessary to keep the 
exact $d$-dimensional coefficient of the double pole terms in 
the one-loop momentum distribution function.

To summarize, when we attempt to interpret (\ref{eq:Pqg1loop}) 
in the SCET framework, we discover two problems with the 
standard treatment of factorization in SCET. First, the 
renormalization and logarithmic terms of some SCET operators 
with singular matching coefficients are not obtained 
correctly. Second, the convolution integrals of the hard matching 
coefficients with the jet functions---the integral over $z$ 
above---diverge and must 
themselves be regularized, for instance dimensionally. 
These obstacles appear first at NLP. 


\subsection{Exponentiation conjecture}

We shall pursue the SCET interpretation further in 
Section~\ref{sec:derivation}. Here, we observe that 
crossing symmetry relates $q \phi^{*}  \to qg$ to $H\to q\bar{q}g$ 
discussed in \cite{Moult:2019uhz}, and we follow 
\cite{Moult:2019uhz} by {\em conjecturing} that the 
leading poles are given correctly to all orders in $\as$ 
by exponentiating the one-loop expression (\ref{eq:Pqg1loop}): 
\bea\label{eq:conjectureqg}
\mathcal{P}_{qg}(s_{qg},z) &=& 
\mathcal{P}_{qg}(s_{qg},z)|_{\rm tree} \,
\exp\Bigg[\frac{\alpha_s}{\pi}\frac{1}{\epsilon^2}\,
\bigg(\T_1\cdot\T_0\left(\frac{\mu^2}{ zQ^2}\right)^{\!\epsilon}
  + \T_2\cdot\T_0\left(\frac{\mu^2}{\bar zQ^2}\right)^{\!\epsilon}
\nn\\
&& + \T_1\cdot\T_2\left(\left(\frac{\mu^2}{Q^2}\right)^{\!\epsilon}
- \left(\frac{\mu^2}{zQ^2}\right)^{\!\epsilon}
+\left(\frac{\mu^2}{zs_{qg}}\right)^{\!\epsilon}\,\right)
\bigg)+{\cal O}\left(\frac{1}{\epsilon}\right)\Bigg]\,,\quad
\eea
where, for the $qg$ case, 
$\T_1\cdot\T_0=C_A/2-C_F$, $\T_2\cdot\T_0=\T_1\cdot\T_2=-C_A/2$, 
but the colour-operator notation is used to emphasize 
the generality of the conjecture.
The exponentiation refers to the $d$-dimensional expression,
since it must be integrated in $d$ dimensions over $z$ as 
discussed above.

In the subsequent section, we employ
consistency relations from the cancellation of poles between
all relevant momentum regions to infer the structure of the
DIS structure function from the contribution with only
hard loops and a single anti-hardcollinear loop,
as shown in the left diagram of Figure~\ref{fig:NLPSCETrep} below. 
The term
involving $s_{qg}$ in the exponent in (\ref{eq:conjectureqg})
arises from the exponentiation of the one-loop
anti-hardcollinear leading pole to all orders,
and should therefore be dropped for this consideration.
Also the tree-level momentum distribution function is non-singular 
as $z\to 1$, hence for the leading poles after the $z$-integration, 
we may replace $\bar z\to 1$. We can therefore simplify
(\ref{eq:conjectureqg}) to
\bea\label{eq:conjectureqghard}
\mathcal{P}_{qg, \rm hard}(s_{qg},z) &=&
\frac{\alpha_sC_F}{2\pi}\frac{1}{z} \,
\exp\Bigg[\frac{\alpha_s}{\pi}\frac{1}{\epsilon^2}\,
\bigg((\T_2\cdot\T_0+ \T_1\cdot\T_2)
\left(\frac{\mu^2}{Q^2}\right)^{\!\epsilon}
\nn\\
&& + \,(\T_1\cdot\T_0-\T_1\cdot\T_2)
\left(\frac{\mu^2}{zQ^2}\right)^{\!\epsilon}\,
\bigg)+{\cal O}\left(\frac{1}{\epsilon}\right)\Bigg]\,,
\eea
where, for the $qg$ case, $\T_2\cdot\T_0+ \T_1\cdot\T_2 = -C_A$
and $\T_1\cdot\T_0-\T_1\cdot\T_2=C_A-C_F$. In SCET we interpret
this as the resummation of the matching coefficient 
of a non-standard B1 operator, squared
and convoluted with the tree-level jet function. 
We shall come back to this in Section~\ref{sec:derivation}, where 
we provide a derivation of this result with factorization 
methods. Notice that the above expression has a homogeneous 
power counting even when $z\ll 1$, since in a $d$-dimensional  
treatment we count $z^{-\epsilon}$
as an $\mathcal{O}(1)$ quantity.

Integrating (\ref{eq:conjectureqghard}) over $z$ yields the 
contribution to the off-diagonal quark DIS structure function 
from any number of hard loops and a single anti-hardcollinear 
loop from the two-particle phase-space, which corresponds to 
the integral over the tree-level final state jet function. 
We obtain
\bea
\label{eq:TphiqVszhard}
W_{\phi,q}\Big|_{q\phi^*\to qg}^{\rm hard} &=& 
\int_0^1 dz\,\left(\frac{\mu^2}{s_{qg}z}\right)^{\!\epsilon}
\mathcal{P}_{qg,\rm hard}\,(s_{qg},z)
\Big|_{s_{qg}=Q^2(1-x)}
\nn\\
&=&  \frac{\alpha_sC_F}{2\pi}\,\left(-\frac{1}{\epsilon}\right) \, 
\left(\frac{\mu^2}{Q^2(1-x)}\right)^{\!\epsilon}
  \exp\bigg[-\frac{\alpha_sC_A}{\pi}\frac{1}{\epsilon^2} \left(\frac{\mu^2}{Q^2}\right)^{\!\epsilon}\bigg]
\nn\\[0.1cm]
&& \times \,\frac{\exp\left[\frac{\alpha_s(C_A-C_F)}{\pi}\frac{1}{\epsilon^2} \left(\frac{\mu^2}{Q^2}\right)^{\!\epsilon}\right]-1}
{\frac{\alpha_s(C_A-C_F)}{\pi}\frac{1}{\epsilon^2} 
\left(\frac{\mu^2}{Q^2}\right)^{\!\epsilon}}
\eea
in the leading-pole approximation. We note that the expression
contains the standard Sudakov factor with colour factor $C_A$ and
a non-standard factor involving the colour factor $C_F-C_A$.
The latter arises from the integral 
\be
\int_0^1 dz\,\frac{\exp[a z^{-\epsilon}]}{z^{1+\epsilon}}
=-\frac{1}{\epsilon}\frac{e^a-1}{a}\,.
\ee

\section{\boldmath 
Resummed off-diagonal partonic cross section and $qg$ splitting function from consistency relations}
\label{sec:consistencyrel}

The theory of deep-inelastic scattering and of resummation 
for $x\to 1$ is based on factorization formulas for the 
scattering of hadrons in terms of partonic quantities. The 
latter are usually IR and ultraviolet (UV) divergent, 
but can be defined 
in terms of a renormalization prescription.  
Consistency relations follow from the requirement that an 
observable must be finite as $\epsilon\to 0$ and allow one
to deduce the expansion in $\epsilon$ of unrenormalized 
partonic quantities based on partial information.

The LL resummation of the quark-gluon splitting function
(\ref{eq:vogtconjecture}) was obtained in \cite{Vogt:2010cv} 
from the requirement that the DIS cross section is an observable 
and hence must be finite, together with additional assumptions 
on the all-order colour structure as well as an exponentiation ansatz
for the full partonic cross section.
A stronger form of consistency relations from pole 
cancellations can be obtained when the regions of virtuality 
relevant to the observable are known. The different 
scaling of every region with the dimensionless parameters 
of the problem implies a larger number of consistency 
relations~\cite{Moult:2016fqy}. For example, 
the LL resummation of the thrust event shapes at NLP was 
derived in \cite{Moult:2018jjd} from the contributions with
a single collinear and an arbitrary number of hard loops
alone and invoking pole cancellations between all regions. 

In this section we use consistency relations to derive the 
NLP LL resummation of the off-diagonal DGLAP kernel 
(\ref{eq:vogtconjecture}) and short-distance coefficient 
function from the exponentiation conjecture (\ref{eq:TphiqVszhard}). 
In this way, we infer the resummation of the short-distance 
functions in the DIS factorization formula from the resummation 
in a single momentum region. In this section we work in moment space, 
following \cite{Vogt:2010cv}, to avoid dealing with convolutions. 
Moments of functions $g(x)$ of the Bjorken scaling variable $x$ 
are taken with the standard definition $g(N)\equiv 
\int_0^1 dx \,x^{N-1} g(x)$. The $x\to 1$ limit corresponds to 
$N\to \infty$ in moment space.

\subsection{Consistency relations for DIS}

In this section, we consider the {\em hadronic} DIS process 
$p+\phi^* \to X$. From standard factorization theorems at 
leading twist in $\Lambda/Q$, where $\Lambda$ denotes the QCD scale, 
we can write the hadronic tensor as 
\begin{equation}
W = \sum_i W_{\phi,i} \,f_i\,,
\label{eq:leadingtwistfact}
\end{equation}
where $i$ sums over all partonic scattering channels,\footnote{In the 
following, we imply the summation convention over repeated 
partonic channel indices and often leave out the sum symbol.}
and $f_i$ denotes the unfactorized (unrenormalized) parton 
distribution function (PDF) of $i$ in the proton $p$. Thus $f_i$ 
contains dimensionally regulated 
UV divergences. The finite,  $\overline{\rm MS}$ subtracted 
parton distributions and short-distance coefficients (partonic 
cross sections) are related to $W_{\phi,i}$, $f_i$ 
by
\begin{equation}
\tilde f_k = Z_{ki} f_i, \qquad W_{\phi,i} = \tilde C_{\phi,k} Z_{ki}\,, 
\label{eq:APzfactor}
\end{equation}
such that 
\be
  W_{\phi,i}f_i=\tilde C_{\phi,k}\tilde f_k\,.
\ee
Note that the short-distance coefficients $\tilde C_{\phi,k}$ 
have finite 
limits for $\epsilon\to 0$, but are still $d$-dimensional.  
The splitting kernels are given by
\be
\label{eq:DGALPkernels}
  P_{ij} = - \gamma_{ij} = \frac{d Z_{ik}}{d\ln\mu}(Z^{-1})_{kj}\,.
\ee

For generic $N$ the leading-twist DIS factorization formula 
involves hard and collinear physics related to the scales 
$Q$ and $\Lambda$. The latter is 
non-perturbative and factorized into the PDFs. 
For large $N$, the small 
invariant mass of the final state (see also
Appendix~\ref{app:largexDIS}) introduces a new scale 
into the problem, which is also the source of the large 
logarithms that we wish to sum. The four relevant virtualities 
are:
\begin{itemize}
\item hard, $p^2=Q^2$
\item anti-hardcollinear, $p^2=Q^2\lambda^2=Q^2/N$
\item collinear, $p^2=\Lambda^2$
\item softcollinear, $p^2=\Lambda^2\lambda^2=\Lambda^2/N$
\end{itemize}
The anti-hardcollinear virtuality arises from the requirement 
of a small-mass final state $X$. In the adopted large-momentum 
frame, its large momentum is in the opposite direction of the 
incoming proton, hence ``anti-hardcollinear''. We also need 
a softcollinear virtuality $\Lambda/N\ll \Lambda$, which 
accounts for the anomalously small momentum of the target 
remnant as $x\to 1$  \cite{Becher:2006mr}. Note, however, that 
there is no soft region in DIS.

The calculation of the DIS process is imagined to be strictly 
factorized into contributions from the different virtualities. 
A multi-loop diagram is considered as a sum of terms, in which 
every loop momentum has one of the above virtualities, in 
the spirit of the strategy of expanding by 
regions \cite{Beneke:1997zp}. Each loop is then associated with 
a factor $(\mu^2/p^2)^\epsilon$ times a function of $\epsilon$, 
which will usually be singular.  
The consistency relations follow from the requirement that 
the sum of all terms is non-singular as $\epsilon\to 0$. 
We note that dimensional regularization only factorizes regions 
with different virtualities. It is not sensitive to the 
scaling of the momentum components separately. However, for 
the present problem, this will be sufficient to obtain 
non-trivial consistency constraints.

\subsubsection{Leading power}

The resummation of leading large-$N$ logarithms at leading power 
is simple 
and well-known. We rederive it here from consistency relations 
and the RGE for the hard 
function to illustrate the method.

At LP  there is no mixing of partonic channels. Only the 
gluon channel contributes to the DIS cross section. We expand the 
diagonal gluon channel in $\alpha_s$ according to
\be
\label{eq:DISLPleadingpole}
W_{\phi,g} \,f_g = 
f_g(\Lambda) \times
  \sum_n \left(\frac{\alpha_s}{4\pi}\right)^n \frac{1}{\epsilon^{2n}}\sum_{k=0}^n\sum_{j=0}^{n} b_{kj}^{(n)}(\epsilon) 
\left(\frac{\mu^{2n}N^j}{Q^{2k}\Lambda^{2(n-k)}}\right)^{\!\epsilon} 
+\mathcal{O}\left(\frac{1}{N}\right)\,.
\ee
Here $k$ denotes the number of hard plus  anti-hardcollinear loops, 
which determines the dependence $Q^{-2k\epsilon}$ on $Q$, 
$j$ is the number of anti-hardcollinear and 
softcollinear loops, which determines the number of times 
the factor $N^\epsilon$ appears. $n-k$ is then the 
number of collinear plus softcollinear loops related to the PDF. 
At $\mathcal{O}(\alpha_s^n)$, the leading singularity is 
$1/\epsilon^{2n}$. This factor has been extracted, such that 
the coefficients $b_{kj}^{(n)}(\epsilon)$ can be  
expanded in non-negative powers of $\epsilon$. For the 
LL resummation, we focus on the leading poles, 
and need only the constant part  $b_{kj}^{(n)} \equiv 
b_{kj}^{(n)}(0)$. At LP, we also drop the $\mathcal{O}(1/N)$ 
correction. The above expansion holds when expressed 
in terms of the dimensionless bare coupling $\mu^{-2\epsilon} 
\alpha_{s0}$, since some of the poles are related to 
coupling renormalization. However, since the relation between the 
bare and renormalized coupling involves at most a single 
pole per loop, we can identify the expansion parameter with 
the renormalized $\overline{\rm MS}$ coupling 
$\alpha_s\equiv \alpha_s(\mu)$ to leading-pole accuracy. 

We regard $f_g$ on the left-hand side as the unrenormalized 
gluon PDF at the factorization scale $\mu$. To make the dependence 
on the collinear and soft-collinear scale explicit, we relate it 
to a non-perturbative reference PDF via
\be
\label{eq:Ugg}
f_g(\mu)=U_{gg}(\mu) f_g(\Lambda) = 
f_g(\Lambda)\,[1+{\cal O}(\alpha_s)]\,,
\ee
where $U_{gg}(\mu)$, defined by this equation, contains the 
evolution from the scale $\Lambda$ to $\mu$. 
Another way of reading (\ref{eq:DISLPleadingpole}) is 
that it represents DIS on a gluon with IR singularities regulated 
non-dimensionally rather than DIS on a hadron. It is only important 
that the left-hand side is finite as $\epsilon\to 0$, so 
all poles on the right-hand side originate from factorization, and 
have to cancel. 

The requirement of pole cancellation implies not only the 
obvious consistency condition
\be
\sum_{k=0}^n\sum_{j=0}^{n} b_{kj}^{(n)}=0\,,
\ee
from the vanishing of the coefficient of the $1/\epsilon^{2n}$ pole. 
In addition, the coefficients of all terms of the form 
\be
 \left(\ln N\right)^r\left(\ln\frac{\Lambda}{Q}\right)^{\!s}
\times \frac{1}{\epsilon^{2n-r-s}}\,,
\ee
must vanish for  $s+r< 2n$, $r,s\geq 0$, since other subleading 
pole terms from the $\epsilon$ expansion of 
$b_{kj}^{(n)}(\epsilon)$ cannot produce the same logarithmically 
enhanced coefficients. This gives the conditions
\be
\sum_{k=0}^n\sum_{j=0}^{n} j^r (n-k)^{s} \,b_{kj}^{(n)}=0\,. 
\ee
Using the binomial expansion of $(n-k)^s$, they are equivalent 
to the consistency relations 
\be
\label{eq:ckjconstraintLP}
\sum_{k=0}^n\sum_{j=0}^{n} j^r k^{s} \,b_{kj}^{(n)}=0 
\qquad\mbox{for}\ s+r< 2n,\ r,s\geq 0 \,.
\ee
At order $\mathcal{O}(\alpha_s^n)$ this provides $2 n^2+n$ 
equations, but only $n^2+2 n$ of them are linearly independent. 
There are $(n+1)^2$ coefficients $b_{kj}^{(n)}$, hence 
at any order in perturbation theory we can determine 
all $b_{kj}^{(n)}$ through consistency equations in 
terms of a single remaining one.

A particularly convenient choice is $b_{n0}^{(n)}$, which 
corresponds to $n$-loop diagrams with only hard loops. We  
show below that $b_{n0}^{(n)}$ can be determined from 
the one-loop hard coefficient  $b_{10}^{(1)}$ by solving a  
RGE, such that the purely 
hard-loop contribution is given to all orders by
\be\label{eq:T2qLPhard}
  W_{\phi,g}^{LP, LL} \Big|_{\rm hard\ loops} = 
\exp\left[-\frac{\alpha_sC_A}{\pi}\frac{1}{\epsilon^2}
\left(\frac{\mu^2}{Q^2}\right)^{\!\epsilon}\,\right]\,,
\ee
which captures the leading-pole part, denoted by ``$LL$'', 
and implies 
\be
  b_{n0}^{(n)} = (-4C_A)^n\,.
\ee
This provides the single condition required at LP to fix all of 
the $b_{kj}^{(n)}$. We make the ansatz
\be
(W_{\phi,g}\,f_g)^{LP, LL} = 
\exp\left[\frac{\alpha_s C_A}{\pi}\frac{1}{\epsilon^2}
\left\{\left(\frac{\mu^2}{Q^2}\right)^\epsilon
-\left(\frac{\mu^2}{\Lambda^2}\right)^\epsilon\right\}(N^\epsilon-1)
\right]\,f_g(\Lambda)\,,
\label{eq:TFsolutionLP}
\ee
which is finite for $\epsilon\to 0$ and contains only products of 
factors of the form $(\mu^2/p^2)^\epsilon$, with $p^2$ any of 
the four relevant virtualities. This corresponds to
\bea
b_{kj}^{(n)} &=& (4C_A)^n \sum_{m_1,m_2,m_3,m_4\geq 0}
\frac{(-1)^{m_2}(-1)^{m_3}}{m_1!m_2!m_3!m_4!}
\delta_{m_1+m_2,k}\delta_{m_3+m_4,n-k}\delta_{m_1+m_3,j}
\nn\\
 &=& (-1)^{k+j}\,(4C_A)^n\sum_{m=m_0}^{{\rm min}(j,k)}
\frac{1}{m!(k-m)!(j-m)!(n-k-j+m)!}\,,
\eea
where $m_0={\rm max}(0,k+j-n)$. We checked (up to $n=10$)  that 
this ansatz indeed satisfies the consistency 
conditions \eqref{eq:ckjconstraintLP}. 
Since the system is fully constrained (equal number of free 
coefficients and linearly independent consistency conditions), this 
solution is the unique solution of the consistency relations 
given (\ref{eq:T2qLPhard}).

Clearly, (\ref{eq:TFsolutionLP}) factorizes into
\bea
W_{\phi,g}^{LP,LL} &=& \exp\left[
\frac{\alpha_s C_A}{\pi}\frac{1}{\epsilon^2}\left(\frac{\mu^2}{Q^2}\right)^\epsilon(N^\epsilon-1)\right]\,,
\label{eq:TgLP}\\
f_{g}^{LP,LL} &=& \exp\left[-\frac{\alpha_s C_A}{\pi}
\frac{1}{\epsilon^2}\left(\frac{\mu^2}{\Lambda^2}\right)^\epsilon(N^\epsilon-1)\right]\,f_g(\Lambda)\,,
\label{eq:fgLP}
\eea
where the first expression is the unfactorized partonic cross 
section, and the second the unfactorized PDF. The latter 
shows that the gluon PDF in the $x\to 1$ limit must be 
considered as a two-scale object already at LP, since 
$f_{g}^{LP,LL}$ depends on the softcollinear in addition to 
the collinear virtuality. Recall that in the derivation of 
these expressions $N^\epsilon$ is treated as an 
$\mathcal{O}(1)$ quantity that must not be expanded in 
$\epsilon$. However, by definition of the $\overline{\rm MS}$ 
scheme, to define the $\overline{\rm MS}$ renormalization 
constants, the pole part is extracted by expanding 
in $\epsilon$ at fixed (large) $N$. From (\ref{eq:fgLP}) 
and the requirement that $\tilde{f}_g$ in (\ref{eq:APzfactor}) be 
finite, we obtain
\bea
Z_{gg}^{LP,LL} &=& \exp\left[\frac{\alpha_s C_A}{\pi}
\frac{\ln N}{\epsilon}\right],
\label{eq:ZMSbarLP}
\\
\tilde C_{\phi,g} &=& \exp\left[\frac{\alpha_s C_A}{\pi}
\frac{1}{\epsilon^2}\left(\left(\frac{\mu^2}{Q^2}
\right)^\epsilon(N^\epsilon-1)-\epsilon\ln N\right)\right]\,.
\label{eq:CMSbarLP}
\eea

The anomalous dimension in the gluon channel is 
obtained from 
\be
\gamma_{gg}(N) = -\left(\frac{d}{d\ln\mu} Z_{gg}\right) 
Z_{gg}^{-1} \,.
\label{eq:gamggabstract}
\ee
In the leading (double) pole approximation, the evolution of 
$d$-dimensional  $\overline{\rm MS}$ coupling is given by 
\be
\label{eq:betafn}
\frac{d\alpha_s}{d\ln\mu} = -2\epsilon \alpha_s\,,
\ee
hence
\be
\gamma_{gg}^{LP,LL}(N) = \frac{\alpha_sC_A}{\pi}\,2\ln N\,,
\label{eq:gamggsolution}
\ee
with no $(\alpha_s \ln^2 N)^k$ corrections. This corresponds 
to the well-known fact that the DGLAP kernel for $x\to 1$ 
at LP is
\be
P_{gg}(x) = \frac{2\Gamma_{\rm cusp}(\alpha_s)}{[1-x]_+} +
2\gamma^g(\alpha_s)\delta(1-x)+\mathcal{O}((1-x)^0)\,,
\ee
with no $\ln^n(1-x)/[1-x]_+$ corrections to the $1/[1-x]_+$ term to 
any order in $\alpha_s$.

\subsubsection{Derivation of the resummed hard function 
(\ref{eq:T2qLPhard})}
\label{sec:T2qLPhard}

In the $x\to1$ limit the QCD part of the Higgs-gluon interaction 
is closely related to the Sudakov form factor for gluon 
scattering. In SCET notation (see Appendix~\ref{app:SCET}), it 
matches to the operator  
\be
J^{A0} = 2 g_{\mu\nu}\,
\nm\partial {\cal A}^{\mu A}_{\perp \overline{hc}}(sn_-)
\np\partial {\cal A}^{\nu A}_{\perp  c}(tn_+)\,,
\label{eq:A0gluoncurrent}
\ee
with a collinear gauge-invariant transverse gluon field in 
the collinear direction of the initial-state gluon and an 
anti-hardcollinear one for the outgoing gluon. The 
square of the hard matching coefficient $C^{A0}$ of this operator 
contains the large logarithms at LP as $x\to 1$ in the DIS 
structure functions, which we associated with 
$W_{\phi,g}^{LP, LL}|_{\rm hard\ loops}$ above.\footnote{See 
Appendix~\ref{app:largexDIS} for a very brief summary of 
factorization for $x\to 1$ at LP.} Here we need the resummation 
of the pole part of the bare coefficient rather than of 
the large logarithms in the renormalized coefficient.

The anomalous dimension of $J^{A0}$ takes the general form 
\be
\Gamma(\alpha_s,\mu) \equiv -\frac{dZ}{d\ln\mu} Z^{-1}
= \Gamma_{\rm cusp}(\alpha_s) \ln\frac{Q^2}{\mu^2} 
+ \gamma(\alpha_s)\,.
\label{eq:cuspAD}
\ee
With 
\be 
\frac{d\alpha_s}{d\ln\mu} \equiv -2\epsilon \alpha_s +\beta(\alpha_s)\,,
\ee
we can solve (\ref{eq:cuspAD}) for \cite{Becher:2009qa} 
\be
\ln Z(\mu) = 
\int_0^{\alpha_s(\mu)} \frac{d\alpha}{\alpha} 
\frac{1}{2\epsilon-\beta(\alpha)/\alpha}
\left(\Gamma(\alpha,\mu) -\int_0^{\alpha}
 \frac{d\alpha'}{\alpha'} 
\frac{2 \Gamma_{\rm cusp}(\alpha')}{2\epsilon-\beta(\alpha')/\alpha'}
\right)\,.
\label{eq''logZ}
\ee
The bare coefficient function is given by 
\be
C^{A0}_{\rm bare} = Z(\mu) C^{A0}(\mu)\,,
\ee
where $C^{A0}(\mu)$ is free of poles. The bare coefficient 
does not depend on $\mu$ and is a function of 
the dimensionless quantities $\alpha_{s0}/Q^{2\epsilon}$ and 
$\epsilon$. The resummation of the pole part is 
obtained most easily by choosing $\mu=Q$, in which 
case $C^{A0}(Q)$ contains no large logarithms, and by 
expressing $\alpha_s(Q)$ in terms of the bare coupling 
$\alpha_{s0}$. 

The cusp anomalous dimension is responsible for the double 
logarithms. To sum the leading poles, the one-loop 
approximation suffices. We therefore set
\be 
\Gamma_{\rm cusp}(\alpha_s) = \frac{\alpha_s C_A}{\pi}\,,
\qquad
\gamma(\alpha_s) =0\,,\qquad
\beta(\alpha_s) = 0\,,
\ee
and obtain 
\be
\ln Z^{LL}(Q) = -\frac{\alpha_s(Q) C_A}{2\pi} 
\frac{1}{\epsilon^2} 
=  -\frac{\alpha_{s0} C_A}{2\pi} 
\frac{1}{\epsilon^2} \frac{1}{Q^{2\epsilon}} 
=  -\frac{\alpha_{s}(\mu) C_A}{2\pi} 
\frac{1}{\epsilon^2} \left(\frac{\mu^2}{Q^2}\right)^{\!\epsilon}\,.
\ee
It is sufficient to use the tree-level coefficient $C^{A0}(Q)=1$ 
to obtain the leading poles. We then find 
\bea
W_{\phi,g}^{LP, LL} \Big|_{\rm hard\ loops} \hspace*{-0.2cm}= 
|C^{A0}_{\rm bare}|^2_{LL} = 
\exp\left(2\ln Z^{LL}(Q)\right) 
= 
\exp\left[-\frac{\alpha_sC_A}{\pi}\frac{1}{\epsilon^2}
\left(\frac{\mu^2}{Q^2}\right)^{\!\epsilon}\,\right]\,,
\qquad
\eea
which proves (\ref{eq:T2qLPhard}).
The above method can be used to include running coupling and 
higher-order effects. However, we restrict ourselves to the 
leading double logarithms here.

\subsubsection{SCET intepretation}

\begin{figure}[t]
\begin{center}
\includegraphics[width=0.30\textwidth]{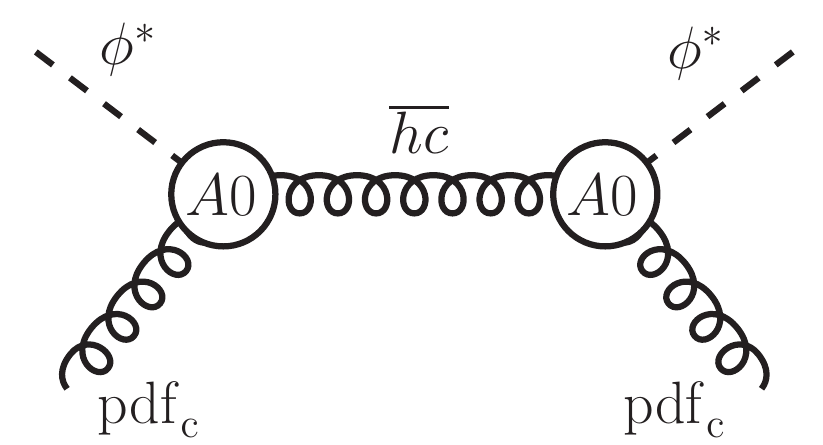}
\end{center} 
\vskip-0.3cm
\caption{Diagrammatic SCET representation of LP resummation. 
The Wilson lines attached to SCET operators are set to 1 
in this graph.}
\label{fig:LPSCETrep}
\end{figure}

The SCET interpretation of DIS at LP as $x\to 1$ is sketched 
in Figure~\ref{fig:LPSCETrep}. A collinear gluon from the 
PDF is converted into an anti-hardcollinear gluon by the 
A0 current (\ref{eq:A0gluoncurrent}), which sources the 
final-state jet. The figure shows the cross section 
$W_{\phi,g} \,f_g$ with hard vertices and lines 
corresponding to the (anti-hard) collinear fields. It does not 
show the Wilson lines attached to these fields. Since 
the softcollinear PDF modes enter at LP only through Wilson 
lines, they do not appear in the graph, despite the fact 
that the result (\ref{eq:fgLP}) for the resummed PDF  
shows that they are necessary at LP to achieve pole 
cancellation. The picture is consistent with the explicit SCET 
computations \cite{Becher:2006mr}. The appearance of 
softcollinear modes only in Wilson lines is also the reason 
why they do not appear explicitly in the LP factorization 
theorem  \cite{Becher:2006mr}, once the standard PDFs 
are introduced. Their presence in (\ref{eq:fgLP}) leads us 
to suspect that this will no longer hold at NLP. There is 
some similarity of this with {\em collinear functions} in 
the factorization of the Drell-Yan process near the kinematic 
threshold. These do not appear in the well-known  LP 
factorization theorem, because soft modes appear only through 
Wilson lines. However, this no longer holds beyond LP, and 
collinear functions do appear at NLP
\cite{Beneke:2018gvs,Beneke:2019oqx}.

\subsection{Next-to-leading power}

Having introduced the method for the well-understood 
case of LP large-$N$ resummation, we proceed to the main subject of 
this paper, the NLP suppressed off-diagonal quark-Higgs 
scattering channel. All partonic channels are relevant at NLP, 
hence we consider the expansion of $\sum_i \,(W_{\phi,i} f_i)$ 
in powers of $1/N$. The LP term $W_{\phi,g}^{LP} f_g^{LP}$ 
was considered before in (\ref{eq:DISLPleadingpole}). The NLP 
term in the hadronic cross section consists of 
\be
\sum_i(W_{\phi,i} f_i)^{NLP} = 
W_{\phi,q}^{NLP} f_q^{LP} + W_{\phi,\bar q}^{NLP}f_{\bar q}^{LP} 
+ W_{\phi,g}^{NLP}  f_g^{LP} + W_{\phi,g}^{LP}  f_g^{NLP}\,.
\ee
The evolution factors that express the unrenormalized PDF at the 
scale $\mu$ in terms of the PDFs at the initial scale 
$\Lambda$ must also be expanded. We generalize (\ref{eq:Ugg}) 
to 
\be
\label{eq:Uij}
f_i(\mu)=U_{ij}(\mu) f_j(\Lambda)\,,
\ee
and find 
\bea
  f_g^{LP}(\mu) &=& U_{gg}^{LP}(\mu)f_g(\Lambda)\,, \nn\\
  f_q^{LP}(\mu) &=& U_{qq}^{LP}(\mu)f_q(\Lambda)\qquad\mbox{
  (similarly for $\bar{q}$)}\,, \nn\\[0.2cm]
  f_g^{NLP}(\mu) &=& U_{gg}^{NLP}(\mu)f_g(\Lambda)+U_{gq}^{NLP}(\mu)
\,(f_q(\Lambda)+f_{\bar q}(\Lambda))\,.
\eea
The LP leading-pole resummed factor
\bea
U_{gg}^{LP,LL}(\mu) &=& \exp\left[-\frac{\alpha_s C_A}{\pi}
\frac{1}{\epsilon^2}\left(\frac{\mu^2}{\Lambda^2}\right)^{\!\epsilon}
(N^\epsilon-1)\right]\,,
\label{eq:UggLP}
\eea
can be inferred from (\ref{eq:fgLP}). $U_{qq}^{LP,LL}(\mu)$ 
and $U_{\bar{q}\bar{q}}^{LP,LL}(\mu)$ are obtained by 
replacing $C_A\to C_F$. The hadronic cross section should be finite 
for any choice of non-perturbative initial conditions 
$f_g(\Lambda)$, $f_q(\Lambda)$ and 
$f_{\bar q}(\Lambda)$. For the off-diagonal quark-gluon 
channel we focus on the terms proportional to $ f_q(\Lambda)$, 
given by 
\be 
\label{eq:Tpgifi}
\sum_i(W_{\phi,i} f_i)^{NLP}\Big|_{\propto f_q(\Lambda)} = 
\left( W_{\phi,q}^{NLP} U_{qq}^{LP} + 
W_{\phi,g}^{LP} U_{gq}^{NLP}\right) f_q(\Lambda)\,.
\ee

\subsubsection{Consistency relations}
\label{sec:NLPconsistency}

Assuming that the same hard, anti-hardcollinear, collinear and 
softcollinear virtualities describe the physics of 
large-$x$ DIS at NLP, we expand the hadronic cross section 
as
\bea
\label{eq:DISNLPleadingpole}
\sum_i (W_{\phi,i} f_i)^{NLP} &=& 
f_q(\Lambda)\times \frac{1}{N}
\sum_{n=1} \left(\frac{\alpha_s}{4\pi}\right)^{\!n} 
\frac{1}{\epsilon^{2n-1}}\sum_{k=0}^n\sum_{j=0}^{n} c_{kj}^{(n)}(\epsilon) \left(\frac{\mu^{2n}N^j}{Q^{2k}\Lambda^{2(n-k)}}\right)^{\!\epsilon} 
\nn\\
&&+ \,f_{\bar q}(\Lambda), \,f_g(\Lambda) \quad\mbox{terms}\,.
\eea
Compared to the previous LP expansion formula 
(\ref{eq:DISNLPleadingpole}), we note the overall NLP factor 
$1/N$, the power $2n-1$ rather than $2n$ for the 
leading pole and the absence of a tree term $n=0$. 
This follows from the fact that a quark must be radiated 
into the final state in the off-diagonal quark-gluon 
channel. This first emission brings a factor of $\alpha_s$ 
but produces only a single $1/\epsilon$ pole. As 
mentioned above, the poles must cancel in all channels 
separately, and we can therefore disregard the 
$f_{\bar q}(\Lambda), \,f_g(\Lambda)$ terms.\footnote{
The $f_g(\Lambda)$ terms can be used to formulate consistency 
relations for the NLP resummation of the gluon channel. 
The antiquark scattering terms are completely analogous to the 
quark terms and need not be considered separately.}
In the following we will only be interested in the leading pole 
at any order, in which case we can replace 
$c_{kj}^{(n)}(\epsilon)$ by their four-dimensional 
values $c_{kj}^{(n)}\equiv c_{kj}^{(n)}(0)$.

The similarity of (\ref{eq:DISLPleadingpole}) and
(\ref{eq:DISNLPleadingpole}) implies that the consistency 
relations from pole cancellation take the same form:
\be
\label{eq:ckjconstraintNLP}
\sum_{k=0}^n\sum_{j=0}^{n} j^r k^{s} c_{kj}^{(n)}=0 
\qquad\mbox{for}\ s+r< 2n-1,\ r,s\geq 0 \,.
\ee
However, the absence of the $1/\epsilon^{2 n}$ pole 
now leads to the condition $s+r< 2n-1$ as compared to  $s+r< 2n$
at LP, (\ref{eq:ckjconstraintLP}). There are still 
$(n+1)^2$ coefficients $c_{kj}^{(n)}$ at order $n$,  
but (\ref{eq:ckjconstraintNLP}) provides only $2 n^2-n$ 
equations. Moreover, not all of them are linearly independent.
We can write $c_{kj}^{(n)}$ as a $(n+1)^2$ dimensional vector 
$c^{(n)}$ in the compound index $[kj]$ with ordering 
$00,01,\ldots 0n, 10, 
\ldots nn$, and regard $j^r k^{s}$ as the entries of a 
$(n+1)^2 \times (2 n^2-n)$ matrix $M^{(n)}$ with indices $[kj]$ and 
$[sr]$ (these ordered as $00, 10, 01,20,11, 02, \ldots$). 
Then (\ref{eq:ckjconstraintNLP}) is expressed 
as $M^{(n)} c^{(n)} =0$. For example, for $n=2$, the $2 n^2-n=6$ 
consistency conditions read in matrix form
\be
\left(
\begin{array}{ccccccccc}
1&1&1&1&1&1&1&1&1\\
0&0&0&1&1&1&2&2&2\\
0&1&2&0&1&2&0&1&2\\
0&0&0&1&1&1&4&4&4\\
0&0&0&0&1&2&0&2&4\\
0&1&4&0&1&4&0&1&4
\end{array}
\right)\cdot
\left(
\begin{array}{c}
c^{(2)}_{00}\\
c^{(2)}_{01}\\
c^{(2)}_{02}\\
c^{(2)}_{10}\\
c^{(2)}_{11}\\
c^{(2)}_{12}\\
c^{(2)}_{20}\\
c^{(2)}_{21}\\
c^{(2)}_{22}
\end{array}
\right)
=0\,.
\label{eq:n2examplefull}
\ee
The number of linearly independent consistency relations is 
related to the rank of matrix $M^{(n)}$, which is $(n+1)^2-3$. 
Hence, the consistency relations allow us to determine 
all $(n+1)^2$ coefficients $c_{kj}^{(n)}$ in terms of 
three unknowns at every order $n$.

\subsubsection{Solution}

Two of the three ``initial conditions'' at every $n$ for 
solving the consistency relations can be fixed trivially. 
In the absence of collinear and softcollinear loops 
($k=n$), there must be at least one anti-hardcollinear 
loop, since the final state cannot be made up of hard 
modes for $x\to 1$. This implies 
\be
  c_{n0}^{(n)}=0\,,
\ee
for all $n$. Similarly, without any hard or anti-hardcollinear 
loops ($k=0$), the necessary off-diagonal $q\to q g$ splitting 
always produces a softcollinear quark. Thus there must be 
at least one softcollinear loop, such that
\be
  c_{00}^{(n)}=0\qquad \mbox{for all $n$\,.}
\ee
The third ``initial condition'' is provided by the Sudakov 
exponentiation conjecture (\ref{eq:TphiqVszhard}). Recall that 
this refers to the all-hard loop corrections to the  
square of the $q\phi^*\to qg$ amplitude integrated over 
the anti-hardcollinear two-particle phase space, which gives the 
series of terms $c_{n1}^{(n)}$ in present notation. 
In moment space, we replace $(1-x) \to 1/N$ in 
(\ref{eq:TphiqVszhard}). Expanding in $\alpha_s$, 
we obtain 
\bea\label{eq:NLPleadingpoleddimDIS}
W_{\phi,q}\Big|^{\rm hard}_{q\phi^*\to qg} &=& 
\sum_{n=1}\left(\frac{\alpha_s}{4\pi}\right)^n\, c_{n1}^{(n)} 
\frac{1}{\epsilon^{2n-1}}
\left(\frac{\mu^{2n}N}{Q^{2n}}\right)^\epsilon\,,
\eea
with 
\be\label{eq:cn1n}
c_{n1}^{(n)} = \frac12 (-4)^n\frac{C_F}{C_F-C_A}
\frac{C_F^n-C_A^n}{n!}=\frac{(-4)^n}{2n!}C_F
\left(C_F^{n-1}+C_F^{n-2}C_A+\dots+C_A^{n-1}\right)\,.
\ee
This particular finite series of $C_F$ and $C_A$ 
terms has already been seen in \cite{Vogt:2010cv,Moult:2019uhz}.

The consistency equations can now be solved. 
For example, at $n=2$, we can rewrite (\ref{eq:n2examplefull}) 
as equations for the unknown coefficients as
\be
\left(
\begin{array}{cccccc}
1&1&1&1&1&1\\
0&0&1&1&1&2\\
1&2&0&1&2&2\\
0&0&1&1&1&4\\
0&0&0&1&2&4\\
1&4&0&1&4&4
\end{array}
\right)\cdot
\left(
\begin{array}{c}
c^{(2)}_{01}\\
c^{(2)}_{02}\\
c^{(2)}_{10}\\
c^{(2)}_{11}\\
c^{(2)}_{12}\\
c^{(2)}_{22}
\end{array}
\right)
=-\left(
\begin{array}{c}
1\\ 2\\ 1\\ 4\\ 2\\ 1
\end{array}
\right)c^{(2)}_{21}\,,
\label{eq:n2examplereduced}
\ee
where we used $c^{(2)}_{00}=c^{(2)}_{20}=0$, and $c^{(2)}_{21}$ is 
given by \eqref{eq:cn1n} for $n=2$. All unknown coefficients are 
uniquely determined, since the matrix has full rank (equal to $6$) 
and is therefore invertible. At general $n$ we can proceed 
analogously, and write the  $n(2n-1)$ consistency conditions in 
the form
\be
   \sum_{k=0}^n\sum_{j=0\atop (jk)\not=(00),(n0),(n1)}^{n} j^r k^{s} c_{kj}^{(n)}=-n^sc_{n1}^{(n)} \qquad\mbox{for}\ s+r< 2n-1,\ r,s\geq 0 \,,
\ee
where the right-hand side is fixed by \eqref{eq:cn1n} and the 
left-hand side can be written as a matrix-vector multiplication 
with a matrix of dimension $((n+1)^2-3) \times n(2n-1)$. This 
matrix is quadratic only for $n=1$ and $n=2$, and has more rows 
than columns for $n\geq 3$. This means the free coefficients are 
over-constrained. Nevertheless, as expected from the previous 
discussion, not all consistency conditions are linearly independent, 
and the rank of the matrix is such that there is a solution, 
which is then the unique solution. 

Rather than attempting a direct solution of these linear 
systems, we will guess a suitable ansatz. From 
(\ref{eq:Tpgifi}), (\ref{eq:DISNLPleadingpole}), we deduce 
that we must match 
\begin{eqnarray}
&&W_{\phi,q}^{NLP} U_{qq}^{LP} + 
W_{\phi,g}^{LP} U_{gq}^{NLP} 
\nonumber\\
&& \stackrel{(\ref{eq:TgLP}),(\ref{eq:UggLP})}{=} 
\,W_{\phi,q}^{NLP} \exp\left[-\frac{\alpha_s C_F}{\pi\epsilon^2}
\left(\frac{\mu^2}{\Lambda^2}\right)^{\!\epsilon}
(N^\epsilon-1)\right] + 
\exp\left[\frac{\alpha_sC_A}{\pi\epsilon^2}
\left(\frac{\mu^2}{Q^2}\right)^{\!\epsilon}(N^\epsilon-1)\right]
U_{gq}^{NLP} 
\nonumber \\
&&\hspace*{0.65cm}\stackrel{!}{=}\frac{1}{N}
\sum_{n=1} \left(\frac{\alpha_s}{4\pi}\right)^{\!n} 
\frac{1}{\epsilon^{2n-1}}\sum_{k=0}^n\sum_{j=0}^{n} c_{kj}^{(n)}
\left(\frac{\mu^{2n}N^j}{Q^{2k}\Lambda^{2(n-k)}}\right)^{\!\epsilon}
\,,
\label{eq:NLPmatch}
\end{eqnarray}
while satisfying (\ref{eq:cn1n}). The form of the LP leading-pole 
solution (\ref{eq:TFsolutionLP}) together with the fact that 
\be
\frac{1}{\epsilon^2}
\left\{\left(\frac{\mu^2}{Q^2}\right)^\epsilon
-\left(\frac{\mu^2}{\Lambda^2}\right)^\epsilon\right\}(N^\epsilon-1)
\ee
appears to be the unique finite combination of all four 
regions in the leading-pole approximation, suggests the 
ansatz 
\begin{eqnarray}
W_{\phi,q}^{NLP} U_{qq}^{LP} + 
W_{\phi,g}^{LP} U_{gq}^{NLP} &=& n(\epsilon)\times
\Bigg\{
\exp\left[\frac{\alpha_s C_F}{\pi}\frac{1}{\epsilon^2}
\left\{\left(\frac{\mu^2}{Q^2}\right)^\epsilon
-\left(\frac{\mu^2}{\Lambda^2}\right)^\epsilon\right\}(N^\epsilon-1)
\right] 
\nn\\
&& \hspace*{-3cm}
-\,\exp\left[\frac{\alpha_s C_A}{\pi}\frac{1}{\epsilon^2}
\left\{\left(\frac{\mu^2}{Q^2}\right)^\epsilon
-\left(\frac{\mu^2}{\Lambda^2}\right)^\epsilon\right\}(N^\epsilon-1)
\right]
\Bigg\}\,,
\label{eq:NLPansatz}
\end{eqnarray}
where $n(\epsilon)$ is a normalization factor yet to be 
determined.\footnote{The relative factor between the two terms 
follows, because there is no $\mathcal{O}(\alpha_s^0)$ term 
in the off-diagonal channel.}
At $\mathcal{O}(\alpha_s)$, (\ref{eq:ckjconstraintNLP}) 
implies $c_{01}^{(1)} + c_{11}^{(1)}=0$, given that $c_{00}^{(1)}=
c_{10}^{(1)}=0$. Expanding (\ref{eq:NLPansatz}) to 
$\mathcal{O}(\alpha_s)$ and matching it to (\ref{eq:NLPmatch})
gives 
\be
n(\epsilon)\,\frac{\alpha_s}{\pi}\frac{C_F-C_A}{\epsilon^2} 
\left\{\left(\frac{\mu^2}{Q^2}\right)^\epsilon
-\left(\frac{\mu^2}{\Lambda^2}\right)^\epsilon\right\}(N^\epsilon-1)
\stackrel{!}{=}
\frac{1}{N}\frac{\alpha_s}{4\pi} \frac{1}{\epsilon} 
c_{11}^{(1)}\left\{\left(\frac{\mu^2}{Q^2}\right)^\epsilon
-\left(\frac{\mu^2}{\Lambda^2}\right)^\epsilon\right\} N^\epsilon\,,
\ee
which yields 
\be
n(\epsilon) = \frac{c_{11}^{(1)}}{4N} \frac{1}{C_F-C_A} 
\frac{\epsilon N^\epsilon}{N^\epsilon-1}
\stackrel{(\ref{eq:cn1n})}{=}
-\frac{1}{2N} \frac{C_F}{C_F-C_A} 
\frac{\epsilon N^\epsilon}{N^\epsilon-1}\,.
\ee
With the normalization determined, (\ref{eq:NLPansatz}) 
reproduces all $c^{(n)}_{n1}$ or (\ref{eq:TphiqVszhard}).
Since (\ref{eq:NLPansatz}) is finite as $\epsilon\to 0$, 
and since the content of consistency relations is 
the finiteness of the physical 
cross section assuming (\ref{eq:TphiqVszhard}), 
(\ref{eq:NLPansatz}) provides the unique solution.
  
Given that $W_{\phi,q}^{NLP}$ must not depend on 
$(\mu^2/\Lambda^2)$, while $U_{gq}^{NLP}$ must not depend on 
$(\mu^2/Q^2)$, the solution (\ref{eq:NLPansatz}) implies 
\bea
W_{\phi,q}^{NLP,LP} &=& -\frac{1}{2N}\frac{C_F}{C_F-C_A}\frac{\epsilon N^\epsilon}{N^\epsilon-1}\Bigg(\exp\left[\frac{\alpha_s C_F}{\pi}\frac{1}{\epsilon^2}\left(\frac{\mu^2}{Q^2}\right)^\epsilon(N^\epsilon-1)\right]\nn\\
  && - \exp\left[\frac{\alpha_s C_A}{\pi}\frac{1}{\epsilon^2}\left(\frac{\mu^2}{Q^2}\right)^\epsilon(N^\epsilon-1)\right]\Bigg)\,,
\label{eq:TphiqNLPsol}\\[0.2cm]
U_{gq}^{NLP,LP} &=& 
-\frac{1}{2N}\frac{C_F}{C_F-C_A}\frac{\epsilon N^\epsilon}{N^\epsilon-1}\Bigg(\exp\left[-\frac{\alpha_s C_F}{\pi}\frac{1}{\epsilon^2}\left(\frac{\mu^2}{\Lambda^2}\right)^\epsilon(N^\epsilon-1)\right]\nn\\
 && -\exp\left[-\frac{\alpha_s C_A}{\pi}\frac{1}{\epsilon^2}\left(\frac{\mu^2}{\Lambda^2}\right)^\epsilon(N^\epsilon-1)\right]\Bigg)\,.
\label{eq:UphiqNLPsol}
\eea
The first of these equations reproduces Eq.~(17) in 
\cite{Vogt:2010cv} for $\mu=Q$ and $C_A= 0$ as assumed there, and 
therefore proves and generalizes the conjectured
all-order structure of the full partonic cross section. 
In addition, the dependence of $W_{\phi,q}^{NLP,LP}$ on $C_F$ and 
$C_A$ is consistent with the colour structure conjectured in Eq.~(13) 
of \cite{Vogt:2010cv}.

It is remarkable that in the leading-pole approximation, the 
full result follows from the exponentiation conjecture for the 
hard-only amplitude (\ref{eq:TphiqVszhard}) by a simple 
substitution. Let us define 
\be 
A \equiv \frac{\alpha_s (C_F-C_A)}{\pi} \frac{1}{\epsilon^2} 
\left(\frac{\mu^2}{Q^2}\right)^{\!\epsilon}, 
\qquad 
S \equiv  \frac{\alpha_s C_A}{\pi} \frac{1}{\epsilon^2} 
\left(\frac{\mu^2}{Q^2}\right)^{\!\epsilon}\,.
\ee
Then the solution of the consistency equations in terms
of the hard-only amplitude 
(\ref{eq:TphiqVszhard}) can be summarized in moment space as 
\bea
\label{eq:NLPsolsubstitute}
&& W_{\phi,q}\Big|_{q\phi^*\to qg}^{\rm hard}=  
\frac{1}{N}\frac{\alpha_sC_F}{2\pi\epsilon}
\left(\frac{\mu^2N}{Q^2}\right)^{\!\epsilon}
  \exp\left[-S\right]
\,\times  \frac{\exp(-A)-1}{A}
\nn\\
&& \longrightarrow \quad
W_{\phi,q}^{NLP,LL} = 
-\frac{1}{N}\frac{\alpha_sC_F}{2\pi\epsilon}
\left(\frac{\mu^2N}{Q^2}\right)^{\!\epsilon}
  \exp\left[S \,(N^\epsilon-1)\right]
\nn\\
&&\hspace*{3.7cm} 
\,\times \,\frac{\exp(A \,(N^\epsilon-1))-1}{A\,(N^\epsilon-1)}\,,
\eea
i.e. $A\to A \,(1-N^\epsilon)$,  $S\to S \,(1-N^\epsilon)$.
The appearance of the factor $(N^\epsilon-1)$ is characteristic 
of the leading-pole solution. The prefactor of the Sudakov 
factors accounts for the anti-hardcollinear $\mathcal{O}(\alpha_s)$ 
contribution that must always be present at NLP.

\subsubsection{DGLAP kernel and coefficient function}

To determine the resummed off-diagonal splitting function and 
the $\overline{\rm MS}$-subtracted short-distance partonic 
cross section, we decompose the
unfactorized partonic cross section $W_{\phi,q}^{NLP,LL}$ 
into its finite and divergent parts. From 
(\ref{eq:APzfactor}) we deduce 
\be
W_{\phi,q}^{NLP}=\tilde C_{\phi,q}^{NLP} Z_{qq}^{LP} + 
\tilde C_{\phi,g}^{LP} Z_{gq}^{NLP}\,,
\label{eq:WNLPinZC}
\ee
where $Z_{qq}^{LP}$ and $C_{\phi,g}^{LP}$ are known from 
(\ref{eq:ZMSbarLP}) (replacing $C_A$ by $C_F$) and 
(\ref{eq:CMSbarLP}), respectively, and the NLP off-diagonal 
factors $ Z_{gq}^{NLP}$, $\tilde C_{\phi,q}^{NLP}$ are to be 
determined. 

From the structure of the LP expressions (\ref{eq:ZMSbarLP}),  
(\ref{eq:CMSbarLP}) it is apparent that the split into 
pole and finite part in the exponents must be done 
according to 
\be
\frac{1}{\epsilon^2}\left(\frac{\mu^2}{Q^2}\right)^\epsilon(N^\epsilon-1) = \frac{\ln N}{\epsilon}+
\frac{1}{\epsilon^2}\left\{\left(\frac{\mu^2}{Q^2}\right)^\epsilon(N^\epsilon-1)-\epsilon \ln N\right\}\,.
\ee 
It will be convenient to introduce the abbreviations
\begin{eqnarray}
&& w\equiv -\epsilon\ln N,\qquad 
a=\frac{\alpha_s}{\pi}(C_F-C_A)\ln^2 N\,,
\\
&&
\widehat{S}_i = 
\frac{\alpha_s C_i}{\pi}\frac{1}{\epsilon^2}
\left\{\left(\frac{\mu^2}{Q^2}\right)^\epsilon(N^\epsilon-1)-\epsilon \ln N\right\}, 
\qquad i=A,F\,,
\end{eqnarray}
which allow us to write (\ref{eq:TphiqNLPsol}) as 
\bea
W_{\phi,q}^{NLP,LP} &=& \frac{1}{2N \ln N}\frac{C_F}{C_F-C_A} 
\,\exp\left[\frac{\alpha_s C_F}{\pi} \frac{\ln N}{\epsilon}\right]
 \frac{w}{e^w-1}\left(e^{a/w} e^{\widehat{S}_A} - e^{\widehat{S}_F}
\right)\,.\quad
\label{eq:newWNLPLLrep}
\eea
Next we note that $w/(e^w-1)$, $e^{\widehat{S}_A}$ and 
$e^{\widehat{S}_F}$ do not have poles in $1/\epsilon$, while 
$\exp\left[\frac{\alpha_s C_F}{\pi} \frac{\ln N}{\epsilon}\right]$ 
matches $Z_{qq}^{LP,LL}$, hence to obtain 
$\tilde C_{\phi,q}^{NLP}$ in (\ref{eq:WNLPinZC}), we must 
separate 
\be 
F(w,a) \equiv \frac{w e^{a/w}}{e^w-1}
\label{eq:Fdef}
\ee 
into its pole and finite part. 
Using 
\be
F(w,0) = \frac{w}{e^w-1} = \sum_{n=0}^\infty \frac{B_n}{n!}w^n\,,
\label{eq:Bernoulli}
\ee
where 
\be B_0 = 1, \quad B_1=-\frac{1}{2}, \quad B_2=\frac{1}{6}, 
\quad B_3=0, \quad B_4 = -\frac{1}{30}, \ldots
\ee
are the Bernoulli numbers, we obtain by expanding 
$e^{a/w}$ that
\bea
F_{\rm pole}(w,a) &=& \sum_{k\geq 1}\frac{1}{w^k}\,
\sum_{n\geq 0}\frac{B_n}{n!(n+k)!}a^{n+k}\,,
\\
F_{\rm fin}(w,a) &=& \sum_{k\geq 0} w^k \sum_{n\geq k}\,
\frac{B_n}{n!(n-k)!}a^{n-k}\,,
\eea
where the sums over $n$ on the right-hand side can be regarded 
as a generalization of Bernoulli polynomials. Inserting this 
decomposition into (\ref{eq:newWNLPLLrep}) and 
matching the resulting expression to (\ref{eq:WNLPinZC}), 
we identify the splitting kernel and short-distance 
coefficient as
\bea\label{eq:Zgq}
Z_{gq}^{NLP,LL} &=& \frac{1}{2N\ln N}\frac{C_F}{C_F-C_A}
\exp\left[\frac{\alpha_s C_F}{\pi}\frac{\ln N}{\epsilon}\right]
\,F_{\rm pole}(w,a)\,,
\\
\tilde C_{\phi,q}^{NLP,LL} &=& \frac{1}{2N\ln N}\frac{C_F}{C_F-C_A}
\Bigg(
F_{\rm fin}(w,a)\exp\left[\frac{\alpha_s C_A}{\pi}\frac{1}{\epsilon^2}\left(\left(\frac{\mu^2}{Q^2}\right)^\epsilon(N^\epsilon-1)-\epsilon\ln N\right)\right]\nn\\
&& -
\frac{w}{e^w-1}\exp\left[\frac{\alpha_s C_F}{\pi}\frac{1}{\epsilon^2}\left(\left(\frac{\mu^2}{Q^2}\right)^\epsilon(N^\epsilon-1)-\epsilon\ln N\right)\right]\bigg)
\,.
\eea
Note that by construction $Z_{gq}^{NLP,LP}$ is a pure pole term, 
hence corresponds to the $\overline{\rm MS}$ PDF renormalization 
factor, while the short-distance coefficient is finite as it 
should be. Indeed, for $\epsilon \to 0$ at fixed $N$, 
which implies $w\to 0$, we find 
\bea
\tilde C_{\phi,q}^{NLP,LL}\Big|_{\epsilon\to 0}
&=& \frac{1}{2N\ln N}\frac{C_F}{C_F-C_A}\Bigg(
{\cal B}_0(a)\exp\left[C_A\frac{\alpha_s}{\pi}\left(\frac12\ln^2 N+\ln N\ln\frac{\mu^2}{Q^2}\right)\right]
\nn\\
&&-\, \exp\left[\frac{\alpha_s C_F}{\pi}\left(\frac12\ln^2 N+\ln N\ln\frac{\mu^2}{Q^2}\right)\right]\Bigg)\,,
\label{eq:Cqgfinite}
\eea
which agrees Eq.~(29) of \cite{Vogt:2010cv} for $\mu=Q$, 
and generalizes it to $\mu\not= Q$. Here
\be
\mathcal{B}_0(a) = F_{\rm fin}(0,a)
\ee
is the Borel transform of the generating function $F(w,0)$ 
of the Bernoulli numbers, already defined in (\ref{eq:B0fn}).

The anomalous dimension in the $q\to gq$ splitting channel is 
obtained from 
\bea
\gamma_{gq}(N) = -\sum_{k=g,q,\bar q}
\left(\frac{d Z_{gk}}{d\ln\mu}\right) 
(Z^{-1})_{kq}  
= \gamma_{gg}^{LP}\,\frac{-Z_{gq}^{NLP}}{Z_{qq}^{LP}}
- \left(\frac{d Z_{gq}^{NLP}}{d\ln\mu}\right) 
(Z^{-1})_{qq}^{LP}\,, 
\label{eq:gamgqabstract}
\eea
where the second equality holds to NLP accuracy and uses 
the vanishing of the off-diagonal terms at LP. Inserting 
the leading-pole resummed results for the $Z$-factors 
and the LP leading-pole anomalous dimension 
(\ref{eq:gamggsolution}) in the gluon-gluon channel, we 
obtain after a short calculation 
\bea
\gamma_{gq}^{NLP,LL}(N) = 
\frac{1}{N}\,\frac{\alpha_s C_F}{\pi}
\left[F_{\rm pole}(w,a) - w\,\frac{d}{da} F_{\rm pole}(w,a)\right]
= - \frac{1}{N}\,\frac{\alpha_s C_F}{\pi} \, \mathcal{B}_0(a)\,,
\label{eq:gamgqsolution}
\eea
which has no poles as it must be and 
proves (\ref{eq:vogtconjecture}) first given in 
\cite{Vogt:2010cv}. We close this derivation with three 
observations:
\begin{itemize}
\item Comparison of the summed large-$N$ anomalous dimensions 
(\ref{eq:gamggsolution}), (\ref{eq:gamgqsolution}) shows that 
there is an infinite series of (double) logarithmic terms 
only for the off-diagonal channel. This implies that not 
only the short-distance coefficient, but also the anomalous 
dimension is a two-scale object in the off-diagonal channel. 
The double logarithms are associated with the colour charge 
change $C_F-C_A$ of the partons that carry large momentum. The 
absence of large logarithms in the diagonal channel seems 
related to the fact that the energetic particles 
have the same colour charge.  
\item The Sudakov exponentiation conjecture was originally 
proposed and explored in \cite{Moult:2019uhz} for the 
case of $e^+ e^-$ or Higgs-decay 
event-shape distributions in the two-jet limit, 
when the final state includes a soft quark. There are 
interesting differences and similarities between 
the DIS and event-shape case, which we elaborate on in 
Appendix~\ref{app:eventshapes}. The solution of the 
consistency relations takes a form similar to 
(\ref{eq:TphiqNLPsol}). The Bernoulli series does not 
arise for event shapes. Event shapes are infrared finite, 
and the Bernoulli numbers arise in DIS from the need 
to factorize the pole part of (\ref{eq:TphiqNLPsol}) 
to obtain the renormalized short-distance coefficient 
and parton distribution, as seen above.
\item Let us comment on similarities and differences compared to the derivation of~\eqref{eq:gamgqsolution} presented in \cite{Vogt:2010cv}.
Both derivations use finiteness and pole cancellation.
In addition, \cite{Vogt:2010cv} conjectures  a specific form of the
full unfactorized partonic cross section (including hard and hardcollinear contributions to all orders)
as well as a particular assumption for the colour structure (as stated in Eqs.~(13) and (14) in \cite{Vogt:2010cv}). We require~\eqref{eq:conjectureqghard}
as an input for the derivation, which involves a single region only (specifically the hard region), that we consider as a weaker assumption compared to those
used in \cite{Vogt:2010cv}. In addition, the exponentiation conjecture~\eqref{eq:conjectureqghard}
lends itself to a derivation based on RGE methods, that we turn to in Section~\ref{sec:derivation}. 
Finally, we obtain
the Bernoulli series in~\eqref{eq:gamgqsolution} by an algebraic derivation in a closed form.
\end{itemize}

\subsubsection{SCET interpretation}

\begin{figure}[t]
\begin{center}
\includegraphics[width=0.30\textwidth]{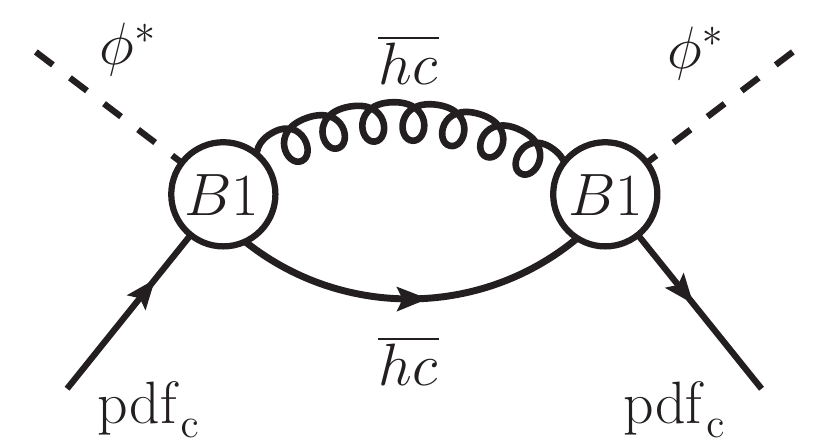}
\hspace*{1.5cm}
\includegraphics[width=0.30\textwidth]{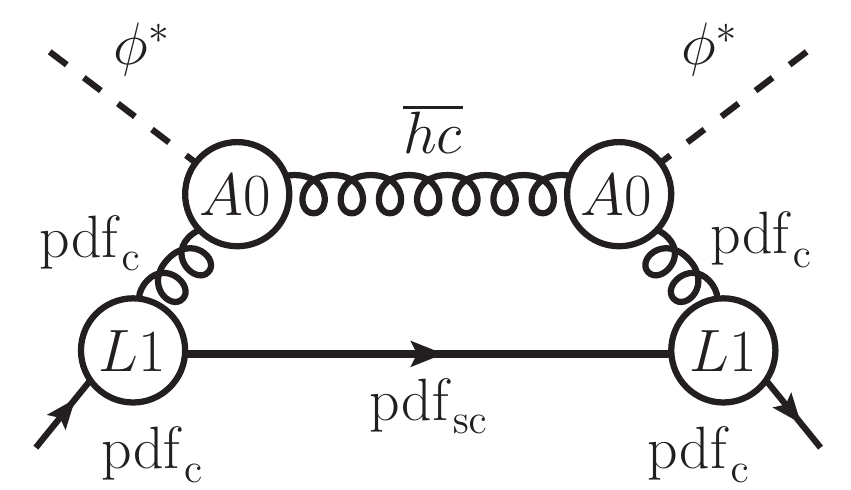}
\end{center} 
\vskip-0.3cm
\caption{SCET representation of the content of  
(\ref{eq:Tpgifi}) for quark-Higgs scattering 
at NLP as $x\to 1$. Wilson lines are set to 1.
\label{fig:NLPSCETrep}
}
\end{figure}

The SCET interpretation of DIS at NLP as $x\to 1$ in the 
off-diagonal channel is sketched 
in Figure~\ref{fig:NLPSCETrep}. The figure shows
\be 
\left(W_{\phi,q}^{NLP} U_{qq}^{LP} + 
W_{\phi,g}^{LP} U_{gq}^{NLP}\right) f_q(\Lambda)
\label{eq:factNLPform}
\ee
from (\ref{eq:Tpgifi}), 
the external quark line representing the quark PDF $f_q(\Lambda)$. 
As before, Wilson lines of whatever fields are set to 
1. 

The left diagram in Figure~\ref{fig:NLPSCETrep} represents 
the first term in this expression and describes the hard 
scattering of a quark off the Higgs boson. The corresponding 
hard vertex is of the B1 type with field content 
$\mathcal{A}_{\,\overline{hc} \perp}
\bar \chi_{\overline{hc}}\chi_{c}$ 
(see Appendix~\ref{app:SCET}), and the presence of two 
anti-hardcollinear fields provides the power suppression. 
The circled operator vertex labelled ``B1'' in the graph 
represents the hard subgraph, which corresponds to 
Figure~\ref{fig:treescattering} at tree level, and to the hard 
region of the diagrams shown in Figure~\ref{fig:oneloopP} 
at one-loop order, respectively. At this 
vertex the incoming quark is converted into the 
anti-hardcollinear quark and gluon in the final state. 
The right diagram describes the second term in 
(\ref{eq:factNLPform}). Here the hard scattering 
occurs through the LP gluon-Higgs scattering 
vertex of the A0 type with field content 
${\cal A}_{\,\overline{hc} \perp} {\cal A}_{\,c \perp}$. 
In the $x\to 1$ limit, the gluon in the $q\to gq$ 
splitting carries almost the entire momentum of the 
initial quark, leaving a remnant softcollinear $q$. 
The interaction that couples soft(collinear) quarks  
to collinear modes is power-suppressed, and part of 
the NLP $\mathcal{L}^{(1)}$ SCET Lagrangian
\cite{Beneke:2002ph,Beneke:2002ni}. In (\ref{eq:factNLPform})
this process is part of $U_{qg}^{NLP}$, the off-diagonal 
evolution of the PDF in the $x\to 1$ limit. Now there 
is a softcollinear mode in the final state, which does 
not arise from a Wilson line, which is shown explicitly 
in the right diagram in Figure~\ref{fig:NLPSCETrep}.

It is evident that both diagrams in the Figure are 
related. If the gluon propagators labelled $\mbox{pdf}_c$ 
in the right diagram became hard, the diagram would 
turn into the left one. What prevents standard SCET 
factorization methods to be applied to this situation, 
is that the convolution of the short-distance coefficient 
of the B1 operator with the final-state jet function is divergent 
after SCET renormalization of the hard and jet 
functions. The above treatment through consistency relations 
circumvents this problem, since it is $d$-dimensional 
to the end. The divergent convolutions do not appear, 
but they are implicit, and done in $d$ dimensions, 
where they exist. In the LL NLP resummation of the diagonal 
quark-quark and gluon-gluon channels for Drell-Yan production 
near threshold with  
SCET methods \cite{Beneke:2018gvs,Beneke:2019mua} these 
complications did not appear, but they do at the 
next-to-leading logarithmic order \cite{Beneke:2019oqx}.

The appearance of an endpoint divergence and the breakdown 
of standard SCET factorization points to the emergence 
of a new scale in the problem, which requires a refactorization 
of the B1-type SCET operator. In the following, we 
show how this idea can be implemented. The resummation of 
logarithms from the new scale will allow us to derive 
the exponentiation conjecture that was used above as a boundary 
condition to solve the consistency equations for 
the NLP LL resummation of the quark-gluon channel.


\section{Derivation of the soft-quark Sudakov factor}
\label{sec:derivation}

In the previous sections, we have seen that the SCET interpretation 
of NLP off-diagonal DIS involves a B1-type current, i.e. an operator 
constructed from one collinear and two anti-hardcollinear SCET 
fields, where the latter two are light-like separated. The operator 
creates the anti-hardcollinear final-state quark and gluon carrying 
momentum fractions $z$ and $1-z$, respectively, of the total 
anti-hardcollinear final-state momentum. The peculiarity of our problem manifests itself in the fact that the convolution of tree-level matching coefficient and the anomalous dimension is not well defined, because the integral exhibits an endpoint divergence as $z\to0$, i.e. when the quark becomes soft. This prevents the standard application of the RGE to the summation of the large logarithms. Instead, we must first consider the limit $z\to 0$ and sum $\ln z$ terms, which become large in the singular region, to all orders, while still working in $d$ dimensions. This goes beyond the standard paradigm of SCET, where the large component of collinear momenta are assumed to be of the order of the hard scale, hence the momentum fraction $z$ appearing in a B-type current is treated as an 
order one parameter $z =\mathcal{O}(1)$. $z$ is nevertheless 
integrated from 0 to 1. This is justified since the 
contribution to the integral from an interval $[0,\delta]$ 
can be made arbitrarily small as $\delta\to0$, as long as the 
matching coefficient is less singular than $1/z$. 

A similar problem has recently been discussed in \cite{Beneke:2019kgv}, where it was noted that the singular part of the hard matching coefficient can be included into the definition of the operator. This leads to the concept of singular and regular B1 operators. In that case, the singular B1 operator was related by reparameterization  invariance (RPI) to the leading-power A0 operator. The singular B1 operator can then be combined with the time-ordered product of the leading-power A0 operator and the NLP Lagrangian to obtain a well-defined operator, whose evolution is governed by the standard RGE. The analogous construction does not work for DIS considered here, as there is no RPI 
relation of the NLP B1 operator to the LP one. 
This fact  is easily  understood from the observation that  the LP and NLP operators relevant to the dicussion of off-diagonal DIS have different fermion numbers in the collinear {\em and} anti-collinear sectors, unlike in \cite{Beneke:2019kgv}, where the NLP current was obtained by adding a gluon building block to the LP current. 

From Figure~\ref{fig:treescattering} it is evident that the 
intermediate gluon propagator is proportional to $1/z$ as $z\to0$. 
When the momentum fraction of the outgoing quark is $\mathcal{O}(1)$, 
the intermediate gluon propagator is hard, and it has to be 
integrated out, which gives the matching coefficient of the B1 
operator that behaves as  $1/z$ for $z\to0$. In this 
limit, however, the virtuality of the intermediate propagator approaches 
zero, and the intermediate gluon should still be present as a 
dynamical mode in the effective theory (EFT) rather than 
having been integrated out. This causes a breakdown of the 
standard application of SCET to this problem, and its failure to 
reproduce the IR singularity structure of full QCD correctly. 
We are therefore forced to revise the structure of the modes in 
the presence of an endpoint-singular matching coefficient.

\subsection{Scales relevant for the 
endpoint-singular B1 operator}

To cure the lack of proper scale separation due to the endpoint 
singularity, we must identify the intermediate scales and modes 
relevant to the  $z\to0$ limit, which goes beyond the SCET$_{\rm I}$ 
paradigm.  Only then, the soft-quark Sudakov expoentiation conjecture~(\ref{eq:conjectureqghard}) can be derived  with EFT methods. 

To understand the scales relevant for our problem, we perform a method-of-region analysis \cite{Beneke:1997zp} of the integrals appearing in the one-loop diagrams shown in Figure~\ref{fig:oneloopP}. We assume that the external momenta are slightly off-shell so that we can identify all modes which contribute to the loop integrals when there is a dimensionful infrared scale. 
As elsewhere in this paper, we focus on the terms giving rise to the leading logarithms. This means that we want to capture terms which diverge in the limit when the quark momentum fraction goes to zero, $z\to0$, and we focus on the double poles in $\epsilon$ or single $\epsilon$ poles multiplied by $\ln z$. 
This allows us to drop many terms and perform simplifications such that the final expression for the leading one-loop correction takes 
the form 
\begin{equation} 
\label{eq:A}
\mathcal{A}= 
{\cal M}^{(1)}_{q \phi^{*}  \to qg} \times 2i\, g_{s}^{2} 
\left(\mathbf{T}_1\cdot \mathbf{T}_0 \,I_{1}
+\mathbf{T}_2\cdot \mathbf{T}_0 \,I_{2}
+\mathbf{T}_1\cdot \mathbf{T}_2 \,I_{12}\right),
\end{equation}
with ${\cal M}^{(1)}_{q \phi^{*}  \to qg}$ 
the Born amplitude for the process $q(p) +\phi^*(q) \to 
q(p_1)+ g(p_2)$. The result can be expressed in terms of the three master integrals
\begin{align}
I_{1}&=2p_1 \cdot p\,\tilde \mu^{2\epsilon}\int\frac{d^{d}k}{\left(2\pi\right)^{d}}\frac{1}{k^{2}}\frac{1}{\left(k-p_{1}\right)^{2}}\frac{1}{\left(k-p\right)^{2}}\,,\\
I_{2}&=2p_2 \cdot p\, \tilde\mu^{2\epsilon}\int\frac{d^{d}k}{\left(2\pi\right)^{d}}\frac{1}{k^{2}}\frac{1}{\left(k-p_{2}\right)^{2}}\frac{1}{\left(k-p\right)^{2}}\,,\\
I_{12}&=2p_2 \cdot p\, \tilde\mu^{2\epsilon}\int\frac{d^{d}k}{\left(2\pi\right)^{d}}\frac{1}{k^2}\frac{1}{\left(k-p_{2}\right)^{2}}\frac{1}{\left(k-p+p_1\right)^{2}}\,.
\end{align}
The integral $I_1$ corresponds to diagram (1) of Figure~\ref{fig:oneloopP}, while the sum of diagrams (2) and (5) can be expressed in terms of the integrals $I_{2}$ and $I_{12}$. The remaining diagrams in Figure~\ref{fig:oneloopP} do not contribute to the terms that we consider here. The modes relevant to the endpoint problem can be identified by expanding these integrals. To perform the expansion by regions we 
introduce the variable 
$z$ as defined 
in \eqref{zDefB} as a 
new power counting parameter and take the limits $1\gg z\gg\lambda$. 
We work to leading order in the $z$ and $\lambda$ expansion. 
We assume the following scaling for the external momenta 
$p_1$, $p_2$, $p$, respectively:
\begin{align}
\label{eq:zsac}
&\text{z-anti-softcollinear} &&	
p_1\sim Q(\lambda^2,\sqrt{z}\lambda,z) && p_1^2\sim z\lambda^2 Q^2\,,
\qquad\\
\label{eq:ahc}
&\text{anti-hardcollinear}&&	
p_2\sim Q(\lambda^2,\lambda,1) && p_2^2\sim\lambda^2 Q^2\,,\\
\label{eq:hc}
&\text{hardcollinear}&&
p\sim Q(1,\lambda,\lambda^2) && p^2\sim \lambda^2 Q^2\,,
\end{align}
with component notation $l\sim (\np l,l_\perp,\nm l)$.
With off-shell external momenta all three integrals are IR and UV 
finite. When the off-shell regulator is set to zero, this computation corresponds to standard one-loop matching of the B1 operator in SCET$_{\rm I}$. Note that we choose the momentum $p$ to have hardcollinear rather than collinear virtuality here to facilitate the interpretation of the result as a SCET$_{\rm I}$ matching computation---only subsequently the hardcollinear modes shall be matched on the collinear modes corresponding to the external 
initial-state momentum in DIS as discussed before.  In the following, we focus only on the pole parts of contributing regions.  

We begin with the $I_{2}$ integral. This is a standard vertex integral exhibiting a double logarithmic enhancement, which can be decomposed into the following loop momentum regions:
\begin{itemize}
\item hard  $k\sim(n_+k,k_\perp,n_-k)\sim Q\left(1,1,1\right)$
\begin{equation}\label{eq:23os}
 \left.I_{2}\right|_{h}	=\frac{i}{16\pi^{2}}\left[-\frac{1}{\epsilon^{2}}+\frac{1}{\epsilon}\ln\frac{Q^{2}}{\mu^{2}}\right]+\mathcal{O}\left(\epsilon^{0}\right)
 \end{equation}

 \item hardcollinear  $k\sim Q\left(1,\lambda,\lambda^{2}\right)$
 \begin{equation}
 \left.I_{2}\right|_{hc}=\frac{i}{16\pi^{2}}\left[\frac{1}{\epsilon^{2}}+\frac{1}{\epsilon}\ln\frac{\mu^{2}}{-p^{2}}\right]+\mathcal{O}(\epsilon^{0})
 \end{equation}
 
  \item anti-hardcollinear  $k\sim Q\left(\lambda^{2},\lambda,1\right)$
 \begin{equation}
 \left.I_{2}\right|_{\overline{hc}}=\frac{i}{16\pi^{2}}\left[\frac{1}{\epsilon^{2}}+\frac{1}{\epsilon}\ln\frac{\mu^{2}}{-p_{2}^{2}}\right]+\mathcal{O}(\epsilon^{0})
 \end{equation}
 
   \item soft $k\sim Q\left(\lambda^{2},\lambda^{2},\lambda^{2}\right)$
 \begin{equation}
 \left.I_{2}\right|_{s}=\frac{i}{16\pi^{2}}\left[-\frac{1}{\epsilon^{2}}-\frac{1}{\epsilon}\ln\frac{Q^{2}\mu^{2}}{\left(-p_{2}^{2}\right)\left(-p^{2}\right)}\right]+\mathcal{O}(\epsilon^{0}) \end{equation}
\end{itemize}
As expected, $ \left.I_{2}\right|_{h}+ \left.I_{2}\right|_{hc}+  \left.I_{2}\right|_{\overline{hc}}+ \left.I_{2}\right|_{s}=\mathcal{O}(\epsilon^{0}) $, and this integral is reproduced by the standard SCET$_{\rm I}$ modes. The soft modes appear here because we assumed that momentum $p$ has hardcollinear virtuality. 
The complete EFT description must take into account that the external momentum $p$ has collinear virtuality, and in this case, the soft modes ought to be replaced by the softcollinear modes. 

The integral $I_{1}$ has a similar mode structure, but in this case 
the hard mode results in a scaleless expression and gives a vanishing 
contribution.  Instead, a new z-hardcollinear mode appears. It is obtained by combining z-anti-softcollinear (\ref{eq:zsac}) and hardcollinear (\ref{eq:hc}) momenta, in analogy with the hard mode being a sum of hardcollinear (\ref{eq:hc}) and anti-hardcollinear~(\ref{eq:ahc}) momenta. 
We find following decomposition of the $I_{1}$ integral into modes:
\begin{itemize}
\item z-hardcollinear $k\sim Q\left(1,z^{1/2},z\right)$
\begin{equation}\label{eq:13os}
 \left.I_{1}\right|_{z-hc}	=\frac{i}{16\pi^{2}}\left[-\frac{1}{\epsilon^{2}}+\frac{1}{\epsilon}\ln\frac{zQ^{2}}{\mu^{2}}\right]+\mathcal{O}(\epsilon^{0})
 \end{equation}

 \item hardcollinear  $k\sim Q\left(1,\lambda,\lambda^{2}\right)$
 \begin{equation}
 \left.I_{1}\right|_{hc}=\frac{i}{16\pi^{2}}\left[\frac{1}{\epsilon^{2}}+\frac{1}{\epsilon}\ln\frac{\mu^{2}}{-p^{2}}\right]+\mathcal{O}(\epsilon^{0})
 \end{equation}
 
  \item z-anti-softcollinear  $k\sim Q\left(\lambda^{2},z^{1/2}\lambda,z\right)$
 \begin{equation}
 \left.I_{1}\right|_{z-\overline{sc}}=\frac{i}{16\pi^{2}}\left[\frac{1}{\epsilon^{2}}+\frac{1}{\epsilon}\ln\frac{\mu^{2}}{-p_{1}^{2}}\right]+\mathcal{O}(\epsilon^{0}) 
 \end{equation}
 
\item soft  $k\sim Q\left(\lambda^{2},\lambda^{2},\lambda^{2}\right)$
 \begin{equation}
 \left.I_{1}\right|_{s}=\frac{i}{16\pi^{2}}\left[-\frac{1}{\epsilon^{2}}-\frac{1}{\epsilon}\ln\frac{zQ^{2}\mu^{2}}{\left(-p_{1}^{2}\right)\left(-p^{2}\right)}\right]+\mathcal{O}(\epsilon^{0}) \end{equation}

\end{itemize}
These results show the emergence of the new scale $\sqrt{z}Q$ related to the endpoint singularity in the $z$-integral (\ref{eq:TphiqVsz}). The new scale is not directly related to the scales present in the factorization of the DIS process at LP. Rather, it is generated dynamically at NLP, due to the breakdown of na\" ive factorization for the B1 current with a singular matching coefficient. As the momentum fraction of a collinear quark becomes parametrically small, the scalar product $p_1 \cdot p$ cannot be treated as being of the same order as $p_2\cdot p$. 

The presence of the new scale manifests itself in a particularly subtle manner for the $I_{12}$ integral. Before we proceed to consider the relevant regions for $I_{12}$, let us look at the on-shell result. For $I_1$ and $I_2$, the on-shell results are equal to the hard and z-hardcollinear contributions, respectively. Thus, they are both 
single scale integrals, though the scale for $I_{1}$ is $zQ^2$, while for $I_{2}$ it is $Q^2$. On the contrary, the on-shell result for $I_{12}$ contains a large logarithm 
\begin{equation}\label{eq:123os}
 \left.I_{12}\right|_{\rm{on-shell}}= \frac{i}{16\pi^{2}} \left[-\frac{1}{\epsilon}\ln z \right]+\mathcal{O}(\epsilon^{0}),
\end{equation}
which cannot be removed by any choice of $\mu$. This observation is troublesome as the on-shell integral is already a two-scale object, and it needs to be factorized to achieve a proper EFT interpretation of the result. Returning to the case of off-shell external momenta, we find contributions from the following integration regions:
\begin{itemize}
\item hard  $k\sim  Q\left(1,1,1\right)$
\begin{equation}
 \left.I_{12}\right|_{h}	=\frac{i}{16\pi^{2}}\left[-\frac{1}{\epsilon^{2}}+\frac{1}{\epsilon}\ln\frac{Q^{2}}{\mu^{2}}\right]+\mathcal{O}(\epsilon^{0})
 \end{equation}
\item z-hardcollinear  $k\sim Q\left(1,z^{1/2},z\right)$
\begin{equation}
 \left.I_{12}\right|_{z-hc}	=\frac{i}{16\pi^{2}}\left[\frac{1}{\epsilon^{2}}-\frac{1}{\epsilon}\ln\frac{zQ^{2}}{\mu^{2}}\right]+\mathcal{O}(\epsilon^{0})
\end{equation}
 \item anti-hardcollinear  $k\sim Q\left(\lambda^{2},\lambda,1\right)$
 \begin{equation}
 \left.I_{12}\right|_{\overline{hc}}=\frac{i}{16\pi^{2}}\left[\frac{1}{\epsilon^{2}}+\frac{1}{\epsilon}\ln\frac{\mu^{2}}{-p_2^2}\right]+\mathcal{O}(\epsilon^{0})
 \end{equation}
 \item z-anti-softcollinear  $k\sim Q\left(\lambda^{2},z^{1/2}\lambda,z\right)$
\begin{equation}
 \left.I_{12}\right|_{z-\overline{sc}}=\frac{i}{16\pi^{2}}\left[-\frac{1}{\epsilon^{2}}-\frac{1}{\epsilon}\ln\frac{\mu^{2}}{-zp_2^2}\right]+\mathcal{O}(\epsilon^{0}) 
 \end{equation}
\end{itemize}
We observe that $ \left.I_{12}\right|_{\rm{on-shell}}= \left.I_{12}\right|_{h}+  \left.I_{12}\right|_{z-hc} $, and the large logarithm appears as the result of a cancellation between the hard and z-hardcollinear contributions. When considering the limit $z\to 0$ it is thus crucial to factorize (\ref{eq:conjectureqghard}) into these two contributions. The cusp anomalous dimension governing the LL resummation should not itself contain large logarithms, which cannot be removed by {\em some} 
choice of the scale $\mu$. If that is not the case, then the anomalous dimension should itself be resummed or factorized.  In the following, we perform a refactorization such that the logarithms from the hard scale may be resummed independently of the logarithms whose origin is the z-hardcollinear scale. Such refactorization is necessary to correctly sum \emph{all} the logarithms of $z$.

\subsection{Resummation}

Having understood the scales relevant to the one-loop result, we attempt to construct the EFT framework to derive~(\ref{eq:conjectureqghard}). At this point, we restrict ourselves to LL accuracy and do not claim that the construction presented here could be used to perform resummation at NLL accuracy or beyond. 
As the analysis of regions showed, we must distinguish the hard scale $Q^2$ and the z-hardcollinear scale $zQ^2$. This suggests 
that the matching of QCD to SCET$_{\rm I}$ should be separated 
into two steps: first we integrate out the hard modes, then we 
the remove z-hardcollinear modes to obtain the EFT at the 
hardcollinear scale $\lambda^2 Q^2$. 

The intermediate gluon propagator in the tree diagram shown in 
Figure~\ref{fig:treescattering} has z-hardcollinear virtuality and 
thus it must be associated with the dynamical degrees of freedom at 
the z-hardcollinear scale. The matching equation at the hard scale 
reads
\begin{equation}
F_{A}^{\mu\nu}F_{\mu\nu}^{A}= C^{A0}(Q^{2},\mu^{2})\,J^{A0}\,,
\end{equation}
where the LP SCET$_{\rm I}$ current
\begin{equation}\label{eq:A0}
J^{A0}=2g^{\mu\nu}n_{-}\partial\mathcal{A}_{\perp\mu}^{A,z-\overline{hc}}
\;n_{+}\partial\mathcal{A}_{\perp\nu}^{A,z-hc}
\end{equation}
represents the point-like coupling of the Higgs boson to two 
gluons in SCET, and $C^{A0}(Q^2,\mu^2)=1$ at tree level. In this theory, the diagram  in Figure~\ref{fig:treescattering}  is represented by the matrix element of the time-ordered product 
\begin{equation}\label{eq:Tproduct}
C^{A0}\left(Q^{2},\mu^{2}\right) \left\langle q(p_{1})g(p_{2})\right|\int d^{d}x\;T\left\{J^{A0},\mathcal{L}_{\xi q_{z-\overline{sc}}}^{\left(1\right)}\left(x\right)\right\}\left|q(p)\right\rangle \,.
\end{equation}
The Lagrangian mediating the z-anti-softcollinear quark interaction with z-hardcollinear modes is
\begin{equation}
\mathcal{L}_{\xi q_{z-\overline{sc}}}^{\left(1\right)}\left(x\right)=\overline{q}_{z-\overline{sc}}\left(x_{-}\right)W_{z-hc}^{\dagger}i\slashed D_{\perp,z-hc}\xi_{z-hc}+h.c.
\end{equation}
Let us note a peculiarity which distinguishes this problem from the one discussed in \cite{Beneke:2019kgv}. The z-anti-softcollinear quark is generated by a subleading-power Lagrangian insertion in the collinear sector. This observation suggests that the endpoint-singular contribution in a collinear sector should be combined with a time-ordered product in the corresponding anti-collinear sector, while in \cite{Beneke:2019kgv}, the time-ordered product and the singular current belong to the same direction. 
The hard matching coefficient $C^{A0}$ of the operator (\ref{eq:A0}) satisfies a standard RGE (\ref{eq:cuspAD}), which at LL accuracy reads
\begin{equation}\label{eq:RGEC}
\frac{d}{d\ln\mu}C^{A0}(Q^2,\mu^2)=
\Gamma_{A0}\,C^{A0}(Q^2,\mu^2) = 
\frac{\alpha_s C_A}{\pi}\ln\frac{Q^{2}}{\mu^{2}} \,C^{A0}(Q^2,\mu^2)\,.
\end{equation}

After this first matching, the SCET with z-hardcollinear modes is 
defined at the $zQ^2$ scale. To describe DIS factorization in the limit $x\to1$, we need an EFT at the scale $\lambda^2Q^2$ where $\lambda^2\sim 1-x$. This EFT should contain only modes with hardcollinear virtuality or lower, as is the case for SCET factorization of DIS at LP.  Besides, we need to include z-anti-softcollinear modes as separate entities represented in the EFT by their own set of fields. The detailed construction of this EFT is left for future work. Fortunately, it is not needed to perform LL resummation, since we already possess all the 
essential ingredients. We discuss the resummation in the 
following, and then provide some partly speculative comments on the SCET with z-anti-softcollinear modes in the following subsection. 

We assume the existence of an operator $J^{B1}$, such that we can 
match the time-ordered product (\ref{eq:Tproduct}) 
\begin{equation}\label{eq:Tmatching}
\int d^{d}x\;T\left\{J^{A0},\mathcal{L}_{\xi q_{z-\overline{sc}}}^{\left(1\right)}\left(x\right)\right\} = D^{B1}(zQ^2,\mu^2)\,J^{B1}
\end{equation}
to this $J^{B1}$, which must be composed of a hardcollinear quark, 
a z-anti-softcollinear quark and an anti-hardcollinear gluon field.
While the matching of the 
z-hardcollinear quark field to the hardcollinear quark field 
is trivial, the non-trivial content of this equation is 
the reinterpretation of the anti-collinear sector, where $zQ^2$ 
now appears as the large scale on which the matching coefficient 
and anomalous dimension of the operator can depend. 
As this operator is supposed to reproduce the IR poles of the QCD 
amplitude, its renormalization factor can be deduced from our previous computation of the amplitude (\ref{eq:A}) in the on-shell limit. 
Using~(\ref{eq:23os}), (\ref{eq:13os}) and (\ref{eq:123os}) we find that the LL UV divergence of this operator can be removed by the counterterm 
\begin{equation}\label{eq:ZB1}
Z_{B1,B1}=1-\frac{\alpha_s}{2\pi}\left[\left(C_{F}-\frac{C_{A}}{2}\right)\left[\frac{1}{\epsilon^{2}}-\frac{1}{\epsilon}\ln\frac{zQ^{2}}{\mu^{2}}\right]+\frac{C_{A}}{2}\left[\frac{1}{\epsilon^{2}}-\frac{1}{\epsilon}\ln\frac{Q^{2}}{\mu^{2}}\right]+\frac{1}{\epsilon}\frac{C_{A}}{2}\ln z\right].
\end{equation}
We implicitly assumed that there is no operator mixing at the LL level 
involved in the matching equation (\ref{eq:Tmatching}), i.e. the complete IR pole in QCD is reproduced in the EFT by $Z_{B1,B1}$. Using
\begin{equation}
\frac{d}{d\ln\mu}C^{A0}\left(Q^{2},\mu^{2}\right)D^{B1}\left(zQ^{2},\mu^{2}\right)J^{B1}=0\,, 
\end{equation}
we find that the matching coefficient obeys at LL accuracy the RGE
\begin{eqnarray}
\label{eq:RGED}
\frac{d}{d\ln\mu}D^{B1}\left(zQ^{2},\mu^{2}\right)&=&
-\left[\left(\frac{d}{d\ln\mu}Z_{B1,B1}\right)Z_{B1,B1}^{-1}+\Gamma_{A0}\right]D^{B1}\left(zQ^{2},\mu^{2}\right)\nn\\ 
&&\hspace*{-2.5cm}=\,\frac{\alpha_s}{\pi}\left(C_{F}-C_{A}\right)\ln\frac{zQ^{2}}{\mu^{2}}\,D^{B1}\left(zQ^{2},\mu^{2}\right)  \equiv\Gamma_{B1}D^{B1}\left(zQ^{2},\mu^{2}\right).
\end{eqnarray}
It is pivotal that $\Gamma_{B1}$ depends only on the scale $zQ^2$.  
The fact that the other logarithms dropped out serves as a 
consistency check of this construction. The anomalous dimension of the $D^{B1}$ coefficient is proportional to $C_{F}-C_{A}$, reflecting earlier observations that double logarithmic enhancement \cite{Soar:2009yh} and endpoint divergence \cite{Moult:2019uhz} vanish in $\mathcal{N}=1$ supersymmetric QCD.

It is straightforward to solve~(\ref{eq:RGEC}) and (\ref{eq:RGED}). 
Recalling that we work with bare, $d$-dimensional objects for the  
boundary terms for the $d$-dimensional consistency relations, we
solve for the bare coefficients, and find 
\begin{align}
\left[C^{A0}\left(Q^{2},\mu^{2}\right)\right]_{\rm bare}&=C^{A0}\left(Q^{2},Q^{2}\right)\exp\left[-\frac{\alpha_s C_A}{2\pi}\frac{1}{\epsilon^{2}}\left(\frac{Q^{2}}{\mu^{2}}\right)^{\!-\epsilon}\right],\nn \\
\left[D^{B1}\left(zQ^{2},\mu^{2}\right)\right]_{\rm bare}&=D^{B1}\left(zQ^{2},zQ^{2}\right)\exp\left[-\frac{\alpha_s}{2\pi}\left(C_{F}-C_{A}\right)\frac{1}{\epsilon^{2}}\left(\frac{zQ^{2}}{\mu^{2}}\right)^{\!-\epsilon}\right].
\label{eq:CDresum}
\end{align}
The product of the square of these two coefficients gives 
(\ref{eq:conjectureqghard}), proving the exponentiation 
of soft-quark Sudakov logarithms conjectured in \cite{Moult:2019uhz}.

\subsection{Tentative EFT interpretation}

While it was not essential to know the exact form of the 
operator $J^{B1}$ to achieve LL resummation, we nonetheless 
comment on the possible structure of SCET with 
z-anti-softcollinear modes. We expect that the operator $J^{B1}$ has the form\footnote{For aestethic reasons we write the Hermitian 
conjugate operators corresponding to antiquark scattering here 
and in  Appendix \ref{app:altB1}.}
\begin{equation}\label{eq:JB1}
J^{B1} =\overline{\chi}_{hc}\gamma_{\mu}\left[in_{-}\partial_{\overline{hc}}\mathcal{A}^{\mu}_{\perp\overline{hc}}\right]\left[\frac{1}{in_{-}\partial_{z-\overline{sc}}}\chi_{z-\overline{sc}}\right].
\end{equation}
In this theory, we must decompose the large component of the 
momentum in the anti-collinear sector into a hardcollinear part, 
which is of the order of $Q$, and a residual momentum of the order 
of $zQ$. Above we accordingly decomposed the anti-hardcollinear 
derivative as $in_{-}\partial=in_{-}\partial_{\overline{hc}}+in_{-}\partial_{z-\overline{sc}} $. 
Unlike in SCET$_{\rm II}$, where the soft modes do not interact with the collinear modes, here the z-anti-softcollinear modes can still interact with the anti-hardcollinear modes. This interaction is responsible for the part of the anomalous dimension proportional to $\mathbf{T}_1\cdot \mathbf{T}_2$. 

To compute the anomalous dimension of (\ref{eq:JB1}), one would need to 
derive the Lagrangian for this z-SCET, which we leave for future 
work.  Instead, extending the notion of singular 
SCET$_{\rm I}$ operators in \cite{Beneke:2019kgv}, we consider 
the family of SCET$_{\rm I}$ operators 
\begin{equation}
J_{B1}^{(n)}=\overline{\chi}_{hc}(0)\gamma_{\mu}\left[(in_{-}\partial)^{1+n\epsilon}\mathcal{A}^{\mu}_{\perp\overline{hc}}\left[
\left(\frac{1}{in_{-}\partial}\right)^{\!1+n\epsilon}
\!\!\!\chi_{\overline{hc}}\right]\right]\!(0),
\end{equation}
which are defined at a single point $x=0$, but contain the 
singular part of the matching coefficient in its definition 
through the inverse derivative. Computing the mixing of 
these operators into themselves together with the assumption that 
the relevant scale is $z Q^2$, we find again the 
product of the resummed coefficient functions in 
(\ref{eq:CDresum}). We provide this alternative derivation 
in Appendix \ref{app:altB1}. In addition, we recover 
the renormalization factor (\ref{eq:ZB1}) as the 
standard $\overline{\rm MS}$ renormalization constant of 
$J_{B1}^{(0)}$, if we expand in $\epsilon$, in which case 
all $J_{B1}^{(n)}$ collapse to $J_{B1}^{(0)}$. 
This gives us confidence that a
 SCET operator whose anomalous dimension is equal to the QCD IR 
poles can in principle be constructed. 

The next step of the EFT construction involves integrating out the hardcollinear scale and matching hardcollinear fields onto the collinear fields which describe the modes inside the PDF. The hierarchy of scales is such that soft modes are also integrated out, and instead soft-collinear modes appear. Similar to the leading power, the anti-hardcollinear fields  in the operator (\ref{eq:JB1}) give rise to the so-called jet function at the level of the amplitude squared. Besides, the cross-section contains a contribution due to the time-ordered product of the LP current and subleading-power soft-collinear Lagrangian, as shown in Figure~\ref{fig:NLPSCETrep}.

\section{Conclusion}
\label{sec:conclusion}

Contrary to the expectation that the resummation of 
large logarithms in $1-x$ in the expansion 
of the off-diagonal parton scattering channels 
in deep-inelastic scattering or Drell-Yan production 
near threshold ought to be simpler than for the 
diagonal channels, since the former vanish
at leading power in the expansion in $1-x$, the 
converse is true. This can already be seen from 
the fact that even the DGLAP splitting functions 
contain an infinite series of double logarithms for 
quark-gluon or gluon-quark transitions, for which 
the formula (\ref{eq:vogtconjecture}) was found 
\cite{Vogt:2010cv} a decade ago, but a method for  
systematic improvements beyond leading logarithms 
is still missing. The difficulty appears to be related to 
the emission of a soft quark rather than gluon in the 
parton splitting, or more 
generally to the change of colour charge of the 
energetic partons in the splitting.

The present work was motivated by the desire to understand 
(\ref{eq:vogtconjecture}) from the perspective of scale 
separation and effective field theory as a necessary step 
towards a general resummation formalism at next-to-leading 
power. In the first part of the paper, we showed that 
given that the relevant modes are hard, collinear, 
soft-collinear and anti-hardcollinear, 
the leading-logarithmic resummation of off-diagonal 
deep-inelastic parton scattering as $x\to 1$ follows from 
the resummation of the purely hard virtual contribution to 
the process. The condition that all $1/\epsilon$ poles in dimensional 
regularization cancel between the various regions is 
sufficient to bootstrap the full solution. For the resummed 
purely hard contribution, which acts as a ``boundary condition'' 
to solve these consistency conditions, we adapted the 
``soft-quark Sudakov'' exponentiation conjecture 
\cite{Moult:2019uhz} from event shapes in $e^+ e^-$ 
collisions to DIS. In this way 
we derived the expression for the resummed off-diagonal 
DGLAP kernel in terms of the series of Bernoulli numbers found 
previously \cite{Vogt:2010cv} directly from algebraic 
all-order expressions, that is, without extrapolating  
the structure of an iteratively generated finite series 
of terms. 

The second part of the paper is concerned with the 
derivation of the ``soft-quark Sudakov'' exponentiation 
of the hard function. The hard function can be alternatively 
interpreted either as the light-cone momentum distribution 
amplitude of the final-state $qg$ pair in the off-diagonal 
$2\to 2$ scattering process, or the matching coefficient of
a B1-type collinear operator in SCET. The crucial 
feature is that the amplitude is singular as $1/z$ when 
the quark momentum fraction $z\to 0$. The failure of 
standard Sudakov resummation or SCET factorization is caused 
by the divergence of the convolution of the hard amplitude 
with the final-state jet function and the emergence of 
the new scale $\sqrt{z} Q$. Based on this observation, we 
derive the previously conjectured 
exponentiation formula through the refactorization of 
certain power-suppressed operators in 
SCET which have endpoint-singular matching 
coefficients. The renormalization group equations 
then exhibit the origin of the peculiar $C_F-C_A$ colour 
factor through an additional exponent related to the 
scale $\sqrt{z} Q$. 

We cannot offer a precise effective field theory 
formulation for this refactorization at this point, 
and, furthermore must note that in this treatment, 
the problem of endpoint-divergent convolutions is side-stepped 
by effectively regulating them dimensionally, since the 
consistency relations and the ``boundary condition'' for their 
solution are formulated in $d$ dimensions for unrenormalized 
objects. A truly satisfactory solution would express the 
result as finite convolutions of properly renormalized 
functions. Nevertheless, we believe that the connection between 
various ideas made manifest here for the first time 
should provide useful insight on NLP resummations. In 
particular, the formalization 
of the refactorization of SCET operators appears to be 
a promising avenue for the systematic understanding of 
resummation at next-to-leading power. 

\subsubsection*{Acknowledgments} 
We would like to thank A. Vogt for correspondence.
This work has been supported by the Bundesministerium f\"ur
Bildung and Forschung (BMBF) grant no.~05H18WOCA1, by 
a Fellini--Fellowship for
Innovation at INFN, funded by the European Union's Horizon
2020 research programme under the Marie Sk\l{}odowska-Curie
Cofund Action, grant agreement no.~754496, and by the Taishan 
scholars program of Shandong province. 
Figures were drawn with {\sc Jaxodraw} \cite{Binosi:2008ig}. 
Calculations were done in part with {\sc FeynCalc} \cite{Shtabovenko:2020gxv}, {\tt FIRE} \cite{Smirnov:2019qkx},
{\sc Package-X} \cite{Patel:2015tea}, 
and {\tt Reduze} \cite{Studerus:2009ye}.

\appendix


\section{SCET conventions and relevant modes}
\label{app:SCET}

\begin{itemize}
\item We define light-like reference vectors $n_\pm$ with 
$n_{\pm}^2=0$, $\np \cdot \nm = 2$. Any four-momentum can 
be decomposed as 
\be \label{Sudakov-Decomp}
p^{\mu} = 
\frac{1}{2} \np p \,\nm^\mu +  \frac{1}{2} \nm p \,\np^\mu + 
p^{\mu}_{\perp}\,.
\ee 
Collinear modes have large $\np p$, anti-collinear modes 
large $\nm p$.
\item SCET operators are conveniently constructed from 
collinear quark and gluon fields, which are invariant under 
collinear gauge transformations 
\be
\chi_{c}(x) = (W_{c}^{\dag} \xi_{c})(x),\qquad
\mathcal{A}_{\perp c}^\mu(x) = W_{c}^\dag(x) 
\, \big[iD_{\perp c}^\mu(x) \, W_{c}(x)\big], 
\label{eq:collinearfields}
\ee
where
\be
\label{eq:Wilsonline}
W_c(x) = \mathbf{P}\exp\left[ig_s\int_{-\infty}^{0}ds\,\np
A_c\left(x+s\np\right)\right]
\ee 
is the collinear Wilson line. Similar definitions apply 
to hardcollinear fields. For anticollinear fields $\np$ and 
$\nm$ are interchanged. 
\item In SCET operators, $J^{Xn}_i$ denotes the product of 
collinear fields for direction $i$, here collinear or 
anticollinear. $X=A,B,\ldots$ refers to the number $1,2,\ldots$ 
of collinear fields from (\ref{eq:collinearfields}), and 
$n$ to the power suppression relative to the 
leading term consisting of a single field 
without derivative \cite{Beneke:2002ph,Beneke:2017ztn}. In 
this paper, when we refer to an A0 or B1 operator, we 
refer to field operators involving a product of 
collinear fields in directions $\np$ and $\nm$ with 
field content ${\cal A}^{\nu}_{\perp\overline{hc}}
{\cal A}^{\mu A}_{\perp c}$, for which we employ the 
short-hand notation $J^{A0}$, and   
$[{\cal A}^{\nu}_{\perp\overline{hc}} 
\bar \chi_{\overline{hc}}] \chi_{c}$, which we refer to 
as $J^{B1}$. Both arise in the matching of the 
Higgs-gluon coupling (\ref{eq:Higgsvertex}) to SCET. 
The explicit forms are
\begin{eqnarray} 
&& \label{eq:A0operator} 
J^{A0}(t,\bar{t}\,) =  
2 g_{\mu\nu} \, n_- \partial {\cal A}^{\nu A}_{\,\overline{hc} \perp}(\bar t n_-)
\, n_+ \partial {\cal A}^{\mu A}_{\,c \perp}(t n_+)\,, \\
&& \label{eq:B1operator} 
J^{B1}(t,\bar{t}_1,\bar{t}_2) =  
\frac{g_{\mu\nu}}{2} \, \Big[ n_- \partial {\cal A}^{\nu 
A}_{\,\overline{hc} \perp}(\bar t_1 n_-) \Big]
\,\bigg[ \bar \chi_{\overline{hc}}(\bar t_2 n_-) 
\frac{\overleftarrow{1}}{i n_- \partial} \bigg]
\,i g_s T^{A} \gamma_{\perp}^{\mu} \, \chi_{c}(t n_+)\,.
\qquad\,\,\,
\end{eqnarray}
The B1 operator as given is the one that appears in 
Section~\ref{sec:consistencyrel}. The one in  
Section~\ref{sec:derivation} looks similar but its 
precise mode content is different and the two B1 operators 
must be carefully distinguished. See Table~\ref{tab:Modes} below 
for a summary of modes and their abbreviations. 
\item The following scaling and power counting variables 
are used in this work:  
$\lambda \sim \sqrt{1-x} \ll 1$
related to factorization DIS at large $x$; 
$\eta \sim  \Lambda/Q\ll \lambda$ 
related to the twist expansion. 
The QCD scale $\Lambda$ appears in modes for 
the non-perturbative PDFs. We consider large-$x$ 
factorization at NLP, but always work at LP in 
the twist expansion parameter $\eta$. For the 
refactorization of the B1 operator in 
Section~\ref{sec:derivation}, we also need 
to consider $z \equiv\nm p_1/\nm (p_1+p_2) \ll 1$, where 
$p_1$ is the momentum of the quark, which becomes soft.

The scalings assigned to the momentum modes 
used throughout the paper are summarized 
in Table~\ref{tab:Modes}.

\begin{table}[t]
	\centering
	\begin{tabular}{|l|c|c|} \hline
 Name & $(n_+ l \, , \, l_{\perp} \,,\, n_- l)$ & virtuality $l^2$ \\ \hline 
hard $[h]$                                          & 
$Q(1\,,\,1\,,\,1)$ & $Q^2$  \\ \hline   
z-hardcollinear                 $[z-hc]$ & 
$Q(1\,,\,\sqrt{z}\,,\, z)$ & $z\,Q^2$  \\
z-anti-hardcollinear $[z-\overline{hc}]$ & 
$Q(z\,,\,\sqrt{z}\,,\,1)$ & $z\,Q^2$  \\
z-soft                            $[z-s]$ & 
$Q(z\,,\,z\,,\,z)$ & $z^2\,Q^2$ \\
z-anti-softcollinear $[z-\overline{sc}]$ & 
$Q(\lambda^2\,,\,\sqrt{z} \, \lambda\,,\,z)$ 
& $z\,\lambda^2 \,Q^2$ \\ \hline
hardcollinear                 $[hc]$ &
$Q(1\,,\,\lambda\,,\,\lambda^2)$ & $\lambda^2 \,Q^2$ \\ 
anti-hardcollinear                 $[\overline{hc}]$ &
$Q(\lambda^2\,,\,\lambda\,,\,1)$ & $\lambda^2 \,Q^2$ \\ 
soft                $[s]$ &
$Q(\lambda^2\,,\,\lambda^2\,,\,\lambda^2)$ & $\lambda^4 \,Q^2$ \\ 
collinear                                       $[c]$ & 
$Q(1\,,\,\eta\,,\,\eta^2)$ & $\eta^2 \,Q^2$ \\
softcollinear                                 $[sc]$ & 
$Q(\lambda^2\,,\,\lambda \, \eta\,,\,\eta^2)$ & $\lambda^2 \,\eta^2 \, Q^2$ \\ \hline
\end{tabular}
\caption{Scaling of the momentum modes relevant for DIS.} 
\label{tab:Modes}
\end{table} 

\end{itemize}

\section{DIS at $x\to 1$}
\label{app:largexDIS}

We briefly summarize some results and definitions for  
DIS off a scalar particle and factorization at 
large $x$ at LP here.

\subsection{DIS off a virtual scalar}

\begin{figure}[t]
\begin{center}
\includegraphics[width=0.36\textwidth]{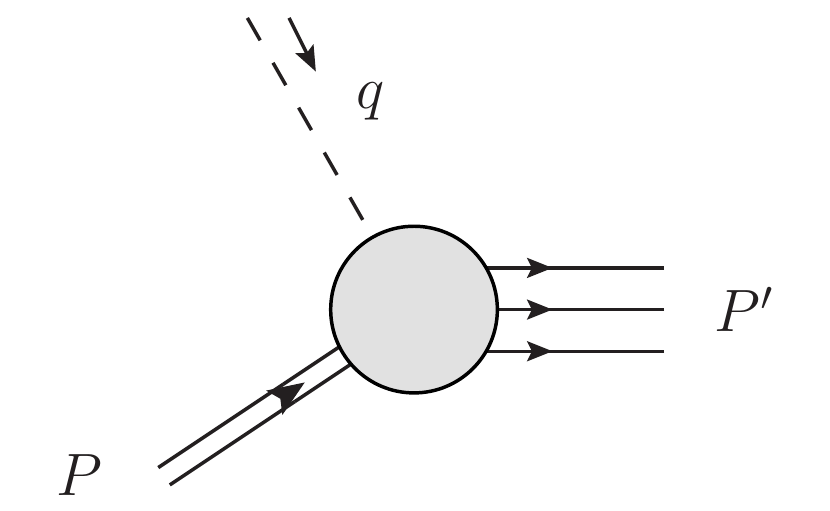}
\end{center}
\caption{DIS process mediated by a scalar boson, as in 
\eqref{DIShadronicgg}.}
	\label{figDISHadr}
\end{figure}

We consider DIS of a particle $N$ off a Higgs boson,
\be\label{DIShadronicgg} 
\phi^{*}(q) + N(P) \to X(P')\,,
\ee
as represented in Figure~\ref{figDISHadr}. 
The large momentum transfer $Q$ and the Bjorken 
scaling variable $x$ are defined by 
\be
Q^2 = -q^2, \qquad \qquad x =   \frac{Q^2}{2 P\cdot q}.
\ee 
Partons in particle $N$ carry momentum 
fraction $\xi$, such that $p = \xi P$, $0<\xi<1$. 
DIS mediated by the exchange of a scalar particle (dubbed 
Higgs boson) occurs via the effective 
gluon-gluon-scalar coupling (\ref{eq:Higgsvertex}),
where the coupling $\kappa$ for an actual Higgs boson 
would be given by
\be\label{kappaDef}
\kappa (m_t,\mu) = \frac{\as(\mu)}{6 \pi v} C_t(m_t,\mu) 
\ee
with
\be
C_t(m_t,\mu) = 1+ \frac{\as(\mu)}{4\pi}
\big(5C_A - 3 C_F \big) + \ord(\as^2).
\ee

The DIS structure function $W_{\phi}$ is defined 
as\footnote{We define $W_{\phi}$ with an additional 
factor of $1/Q^2$ compared to the more common definition 
for DIS of an off-shell photon, to compensate for the 
dimensionful coupling $\kappa \sim 1/v$. An average over 
the spin and colour of the state $N(P)$ is implicitly 
understood when taking the matrix element.} 
\be \label{hadrtensdefgg}
W_{\phi} = \frac{1}{8\pi Q^2} \int d^4 x \, e^{i q \cdot x} \,
\big\langle N(P) \big| \big[G_{\mu\nu}^A G^{\mu\nu A}\big](x)
\big[G_{\rho\sigma}^B G^{\rho\sigma B} \big](0) 
\big| N(P) \big\rangle\,.
\ee
The QCD factorization theorem relates the  
structure function $W_{\phi}$ to the 
partonic partonic short-distance coefficients  
$\tilde C_{\phi,i}(x)$
by means of the convolution 
\be\label{Wpartonicgg}
W_{\phi}(x) = \int_x^1 \frac{d \xi}{\xi} \, 
\bigg\{\tilde{f}_{g}\bigg(\frac{x}{\xi}\bigg)\, 
\tilde{C}_{\phi,g}(\xi)
+ \sum_q \, \tilde{f}_{q} \bigg(\frac{x}{\xi}\bigg)\, 
\tilde{C}_{\phi,q}(\xi)\bigg\}
\ee
with renormalized PDFs $\tilde{f}_i$ of parton $i$ in $N$.
The notation follows \eqref{eq:leadingtwistfact}, 
\eqref{eq:APzfactor} of the main text. There we often 
refer to unfactorized (unrenormalized) partonic structure functions 
$W_{\phi,i}(x)$ and PDFs $f_i$, related to the above 
by \eqref{eq:APzfactor}. Eq.~\eqref{Wpartonicgg} also holds 
with $\tilde{f}_i\otimes\tilde C_{\phi,i} \to 
f_i \otimes W_{\phi,i}$. The equation also applies to 
the case when the particle $N$ is itself a quark or gluon, 
but then the left-hand side is IR divergent for 
massless, on-shell partons. When the IR divergences 
are regulated non-dimensionally with a regulator introducing 
the scale $\Lambda$, the consistency arguments based on 
pole cancellations used in the main text apply to this 
partonic scattering. If dimensional regularization is 
employed, the  unfactorized partonic structure functions 
$W_{\phi,i}(x)$ do not change but the unrenormalized 
PDFs $f_i(x)$ become trivial, because the loop integrals 
are scaleless, and the left-hand side of~\eqref{Wpartonicgg} 
is simply $W_{\phi,i}$. 

\begin{figure}[t]
\begin{center}
\includegraphics[width=0.80\textwidth]{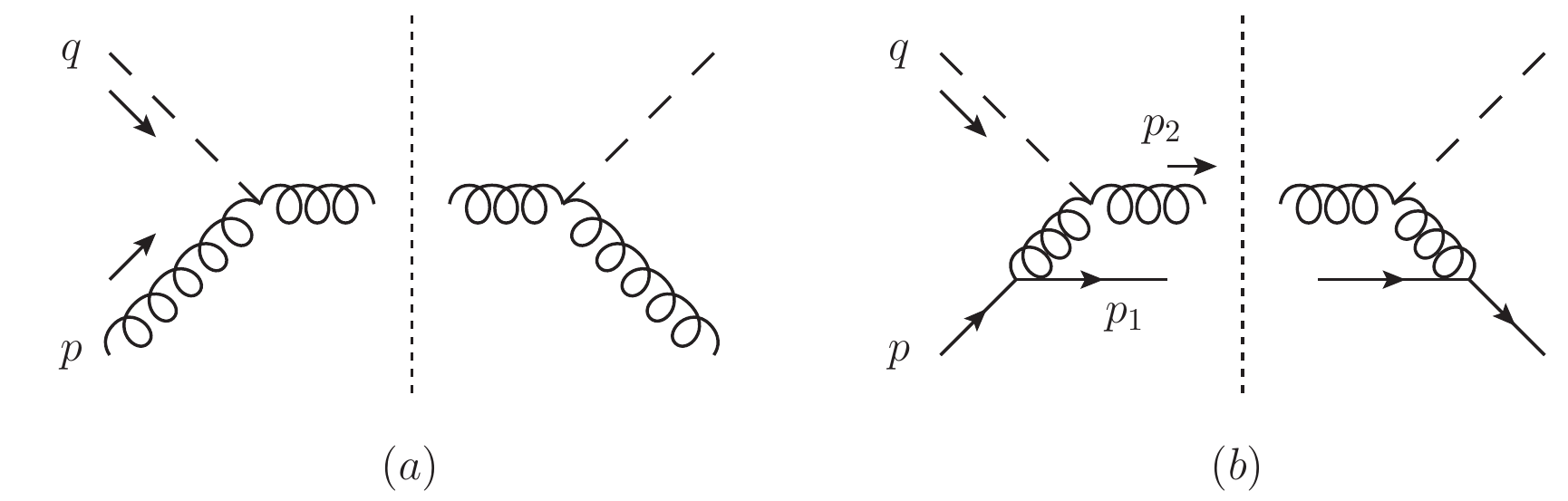}
\vspace*{-0.2cm}
\end{center}
\caption{Partonic contribution to DIS with scalar exchange: (a)
LO diagram; (b) NLO contribution in the quark-gluon channel.}
\label{FigPartonic}
\end{figure}

The hadronic structure function is related to the
phase-space integrated, initial-state spin- and colour-averaged 
and final-state spin- and colour-summed scattering amplitude 
as 
\be\label{MsqHadr}
\int d \Phi_X \, |{\cal M}_{N \phi^{*} \to X}|^2 
= 2\pi \kappa^2 Q^2\, W_{\phi}\,.
\ee
The same relation applies to the partonic 
structure functions for the scattering of gluons and quarks. 
The respective lowest order 
contributions in powers of the 
strong coupling are obtained from 
\bea
\label{partonicWdefgg} 
\int d\Phi_1
\big| {\cal M}_{\phi^{*} g \to g}\big|_{\rm tree}^{2} &=& 
2\pi \kappa^2 Q^2 \,  W_{\phi,g}\big|_{\ord (\as^0)}, \\
\label{partonicWdefqg} 
\int d\Phi_2 
\big| {\cal M}_{\phi^{*} q \to qg} \big|_{\rm tree}^{2} &=& 
2\pi \kappa^2 Q^2\,   W_{\phi,q}\big|_{\ord (\as)}\,.
\eea
The  two-particle phase space $d\Phi_2$ has
been defined in \eqref{PS2def}, and $d\Phi_1$ denotes 
the $d$-dimensional one-particle phase space. 
The tree-level contribution  (see diagram (a) in 
Figure \ref{FigPartonic}) for gluon scattering is
\be\label{MggTree}
 \big| {\cal M}_{\phi^{*} g \to g} \big|_{\rm tree}^{2} 
 = 4 \kappa^2  (1-\eps) \, (p \cdot q)^2\,,
\ee
which inserted in \eqref{partonicWdefgg} gives
\be\label{WphiTreeB}
W_{\phi,g}(x) = \frac{1-\eps}{x} \,\delta(1-x) + \ord(\as)\,.
\ee
The quark-scattering channel starts contributing at $\ord(\as)$ 
with diagram (b) in Figure \ref{FigPartonic}. The 
spin- and colour-averaged / summed matrix
element squared expressed in terms of the variable $z$ 
defined in \eqref{zDefB} reads 
\be\label{Mqg1}
\big| {\cal M}_{\phi^{*} q \to qg} \big|^{2} = 
4 \kappa^2 g_s^2 C_F\, \frac{1-\eps}{2} 
\, \frac{Q^2}{x} \frac{\bar z^2}{z} + \ord(\as^2)\,.  
\ee
Inserting this expression into \eqref{partonicWdefqg},
we get 
\bea
\label{Wsq1Full} 
\nn 
W_{\phi,q}(x)\big|_{\ord (\as)} &=& - \frac{\as C_F}{2\pi} 
\frac{1}{x} \bigg(\frac{\mu^2}{s_{qg}}\bigg)^{\!\eps}\, 
\frac{(1-\eps)\left(1-\frac{\eps}{2}\right)}{\eps(1-2\eps)}
\frac{e^{\gamma_{E} \eps}\,\Gamma(1-\eps)}{\Gamma(1-2\eps)}  \\ 
&=& \frac{\as C_F}{2\pi} 
\bigg[ -\frac{1}{\eps}-\frac{1}{2} -\ln \bigg( \frac{\mu^2}{Q^2(1-x)}\bigg) 
+ \ord(\eps) \bigg] + \ord(\lambda^2)\,,\quad
\eea
with  $s_{qg}$ as defined in (\ref{eq:sqg}).

\subsection{\boldmath Large $x$}

In this paper we focus on the threshold region $x\to 1$. 
This region is characterized by the fact that the 
scattering leaves a soft target nucleon (or parton, depending 
on whether we consider the hadronic or partonic threshold) 
and a jet-like final state $X$ with parametrically small invariant 
mass squared $p^2_X = Q^2 (1-x)/x \ll Q^2$. 
The existence of two scales for $x \to 1$ is the basis of 
the factorization of the partonic structure functions 
$W_{\phi,i}(x)$ into a hard and jet 
function \cite{Sterman:1986aj,Catani:1989ne,Korchemsky:1992xv}. 
For the structure function $W_{\phi}$, focusing on the 
leading gluon scattering channel, such factorization
takes the form 
\be\label{Wfact}
W_{\phi}(x) = H(Q^2,\mu) \int_x^1 \frac{d \xi}{\xi} \, 
J\bigg(Q^2 \frac{1-\xi}{\xi},\mu \bigg) \,
\frac{x}{\xi}f_{g}\bigg(\frac{x}{\xi},\mu\bigg),
\ee
valid to leading power in $\lambda \sim (1-x)$ and in 
$\eta \sim \Lambda/Q$. 
We assume $\lambda \gg \eta$, in which case  
the hard and jet functions in \eqref{Wfact} can be 
calculated in perturbation theory. A similar equation 
holds for the {\em partonic} structure 
functions  $W_{\phi,i}(x)$ themselves, as discussed above, 
if we interpret $f_g$ as the distribution of gluons in parton $i$.

The factorization theorem \eqref{Wfact} can also be derived  
within SCET \cite{Becher:2006mr}. Near threshold the parton 
undergoing the hard scattering has collinear momentum scaling 
$p = \xi P \sim Q(1, 0, \eta^2)$  and carries away 
almost all momentum of the initial state $N(p)$. The 
target remnant is then made out of partons which must have 
softcollinear momenta, i.e. $p_{\rm remnant} \sim Q(\lambda^2,
\lambda \eta,\eta^2)$. This explains the need for softcollinear 
modes in the PDFs, as listed in Table~\ref{tab:Modes}.  
The energetic parton scattering off the Higgs boson (or 
virtual photon in standard DIS) is converted into a jet 
of partons with anti-hardcollinear momentum, 
$p_{\overline{hc}}\sim Q(\lambda^2,\lambda,1)$. 
The factorization theorem \eqref{Wfact} is derived 
by constructing the sequence  
\be\label{EFTtower}
\mbox{QCD} \quad \to \quad \mbox{SCET}(\overline{hc},c,sc)
 \quad \to \quad \mbox{SCET}(c,sc)
\ee
of effective theories, where in the first matching step 
the hard modes are integrated out, and the 
QCD gluon-gluon-scalar interaction is matched onto
A0, B1 etc. SCET($\overline{hc},c,sc$) 
operators built from the collinear gauge-invariant 
building blocks (see Appendix~\ref{app:SCET}). In the second 
step, the jet function emerges as the matching coefficient 
containing the anti-hardcollinear final-state, leaving 
a parton distribution made up of collinear modes and the 
softcollinear target remnant modes.
At LP, only an A0-type current is required, and the hard 
function is given by the square of its short-distance 
coefficient, 
\be
H(Q^2,\mu) = |C^{A0}(Q^2,\mu)|^2\,.
\ee
More important for the present work is the observation 
\cite{Becher:2006mr} that the softcollinear mode appears 
only through a Wilson line in the definition of the 
PDF for $x\to 1$. It is then possible to identify this PDF 
with the standard PDF. The two-step matching 
scheme \eqref{EFTtower} should be expected to hold beyond 
the LP.  However, when \eqref{Wfact} is naively generalized 
to NLP, the convolutions of generalized 
renormalized hard and jet functions diverge. As discussed in 
the main text, this requires, at least for the present, 
a partly $d$-dimensional treatment and a refactorization 
within $\mbox{SCET}(\overline{hc},c,sc)$ to generate the 
correct large-$x$ logarithms that would otherwise be missed. 


\section{Alternative derivation of the LP 
solution (\ref{eq:TFsolutionLP})}
\label{app:alt3.12}

There is a simpler way to obtain the leading-power leading-pole 
expression  (\ref{eq:TFsolutionLP}) for
$(W_{\phi,g}\,f_g)^{LP, LL}$, which bypasses the combinatorially  
involved solution for the coefficients $b_{kj}^{(n)}$ in 
the consistency relation. 

At any $N$ the DIS factorization theorem implies multiplicative 
factorization. At LP in $1/N$, only the gluon channel 
contributes. We can therefore write the expansion for 
the logarithm of the DIS cross section  
(\ref{eq:DISLPleadingpole}) as 
\bea
\label{eq:DISLPleadingpoleLOG}
\ln \left(W_{\phi,g} \,f_g\right) &=& 
\sum_{n=1} \left(\frac{\alpha_s}{4\pi}\right)^{\!n} 
\frac{1}{\epsilon^{n+1}}\sum_{j=0}^{n} 
\left[t_{j}^{(n)}(\epsilon) 
\left(\frac{\mu^{2n}N^j}{Q^{2n}}\right)^{\!\epsilon} 
+ f_{j}^{(n)}(\epsilon)
\left(\frac{\mu^{2n}N^j}{\Lambda^{2n}}\right)^{\!\epsilon} 
\right] 
\nonumber
\\ &&+\,\ln f_g(\Lambda) +\mathcal{O}\left(\frac{1}{N}\right)\,,
\eea
using that $W_{\phi,g}$ cannot depend on $\Lambda$, and 
$f_g$ cannot depend on $Q$. The form of this expansion 
contains the non-trivial statement that for the logarithm 
of the cross section the highest 
pole at $\mathcal{O}(\alpha_s^n)$ is $1/\epsilon^{n+1}$. 

For the leading poles, we can drop the $\epsilon$-dependence 
of the coefficients $t_{j}^{(n)}(\epsilon)$, $ f_{j}^{(n)}(\epsilon)$.
Instead of the $(n+1)^2$ coefficients $b^{(n)}_{kj}$ at 
$\mathcal{O}(\alpha_s^n)$, 
we now have only $2(n+1)$. There are $2 n+1$ consistency 
relations from the requirement of pole cancellation, 
\bea
&& \sum_{j=0}\,j^r (t_{j}^{(n)} + f_{j}^{(n)}) = 0 
\qquad\mbox{for}\qquad r=0,\ldots, n\,,
\\
&& \sum_{j=0}\,j^r f_{j}^{(n)} = 0 
\qquad\mbox{for}\qquad r=0,\ldots, n-1\,,
\eea
leaving one undetermined coefficient per order $n$, which we 
choose to be $t_0^{(n)}$. The solution 
to the consistency relations is 
\be 
t_j^{(n)} = (-1)^j \frac{n!}{j! (n-j)!}\,t_0^{(n)}, 
\qquad f_j^{(n)} = - t_j^{(n)}\,,
\ee
resulting in 
\be
\ln W_{\phi,g}^{LP,LL} = \sum_{n=1} 
\left(\frac{\alpha_s}{4\pi}\right)^{\!n} 
\frac{1}{\epsilon^{n+1}}\,(-1)^n
\left(\frac{\mu^2}{Q^2}\right)^{\!n\epsilon}
(N^\epsilon-1)^n\,t_0^{(n)}\,,
\label{eq:LPsolutionfromlogs}
\ee
and a corresponding expression for $\ln f_g^{LP,LL}$ with 
$Q\to \Lambda$ and an overall minus sign. Comparison with 
the resummed hard-only expression (\ref{eq:T2qLPhard}) 
implies that the logarithm of $W_{\phi,g}^{LP,LL}$ is 
one-loop exact, that is $t_0^{(1)}=-4 C_A$ and $t_0^{(n)}=0$, 
$n>1$. This conclusion can also be reached directly, 
without the explicit resummed result for $W_{\phi,g}^{LP,LL}$, 
from the requirement that 
$\gamma_{gg}^{LP}(N)$ should have at most a single power of $\ln N$ 
at any order in $\alpha_s$, and the observation that 
$\gamma_{gg}^{LP}(N)$ is related to the coefficient of the 
single pole in (\ref{eq:LPsolutionfromlogs}). We then 
recover the previous results (\ref{eq:TgLP}), (\ref{eq:fgLP}).


\section{Alternative derivation of the 
resummed singular B1 current}
\label{app:altB1}

In this Appendix we present an alternative derivation 
of the exponentiation conjecture \eqref{eq:conjectureqghard} for 
the momentum distribution $\mathcal{P}_{qg,\rm hard}(s_{qg},z)$ in the limit $z\to 0$
that is complementary to the presentation in Section~\ref{sec:derivation}. Both rely on the observation
made in the first part of Section~\ref{sec:derivation} concerning the relevant regions for $z\to 0$. However,
instead of considering a ``refactorization'' into hard and z-hardcollinear regions as in Section~\ref{sec:derivation},
the derivation presented here solely relies on the SCET$_{\rm I}$ description of DIS for large $x$ involving
collinear, anti-hardcollinear and softcollinear modes.

We start from the SCET$_{\rm I}$ description of the $q\phi^*\to qg$ scattering process depicted in Figure~\ref{fig:treescattering}
assuming that the incoming $q$ has collinear scaling, and the outgoing $q$ and $g$ are both anti-hardcollinear.
The relevant SCET B1 current has field content $\bar\chi_{c}\,{\cal A}_{\overline{hc},\perp}\chi_{\overline{hc}}$, and
its tree-level matching coefficient diverges as $1/z$ for $z\to 0$, where $z$ is the momentum fraction of the anti-hardcollinear
quark. The momentum distribution $\mathcal{P}_{qg,\rm hard}(s_{qg},z)$ is related to the square of the B1 matching coefficient (note
that one factor of $1/z$ cancels when computing the matrix element squared for $q\phi^*\to qg$). 

As a first attempt one may naively apply RG evolution to the B1 matching coefficient. Keeping only the cusp
part of its anomalous dimension (that is diagonal in Lorentz and spinor indices as well as with respect to the momentum fraction)
\be\label{eq:Bqcuspnonsing}
  \Gamma_{B1,{\rm non-singular}}^{\rm cusp} = \frac{\alpha_s}{\pi}\left( \T_1\cdot \T_0\ln\frac{\mu^2}{zQ^2}+\T_2\cdot \T_0\ln\frac{\mu^2}{\bar zQ^2}\right)\,,
\ee
and performing a $d$-dimensional RG evolution analogously as described in Section~\ref{sec:T2qLPhard} 
yields precisely the exponential factors involving $\T_1\cdot\T_0$ and $\T_2\cdot\T_0$ in \eqref{eq:conjectureqghard} (for $ z\to 0$, i.e. setting $\bar z\to 1$), 
but misses the contribution involving $\T_1\cdot\T_2$. The reason is that the one-loop anomalous dimension for the B1 operator with generic momentum
fractions of the anti-hardcollinear quark and gluon \cite{Beneke:2017ztn}
contains contributions that would diverge in four dimensions, when convoluted with a matching coefficient that goes like $1/z$.

To avoid the problem of divergent convolutions, we retain a $d$-dimensional description, and
define a singular B1 current analogous to \cite{Beneke:2019kgv}, that absorbs the $1/z$ factor 
\begin{equation}
J_{B1}^{\left(0\right)}=\overline{\chi}_{c}(0)\gamma_{\mu}\left[in_{-}\partial\mathcal{A}^{\mu}_{\perp\overline{hc}}\left[\frac{1}{in_{-}\partial}\chi_{\overline{hc}}\right]\right]\!(0)\,.
\end{equation}
This operator essentially agrees with the one considered in 
Section~\ref{sec:derivation}. However, in the present discussion, 
we do not consider a further refactorization. 
Also note that, in contrast to the singular B1 current considered in \cite{Beneke:2019kgv}, the inverse derivative acts on the quark building block instead of the
gluon building block. For the latter case, the matching coefficient of the singular B1 current was linked to the leading power coefficient by reparameterization
invariance, such that its anomalous dimension coincides with the one of the corresponding LP current. 
This relation does not exist in the present case. 

In order to find the anomalous dimension of $J_{B1}^{\left(0\right)}$, we compute its off-shell regulated one-loop matrix element in an external state with
momentum fraction $z$ of the anti-hardcollinear quark. We are interested
in the double-pole part in the limit $z\to 0$, when counting factors of $z^\epsilon$ as order one. Apart from the expected cusp part in accordance with
\eqref{eq:Bqcuspnonsing}, one finds an additional piece involving a factor $\alpha_s\,\T_1\cdot\T_2\,(z^{-\epsilon}-1)/z$. When expanding for $\epsilon\to 0$
this gives a factor $1/z\times \ln z$ that cannot be interpreted as a renormalization of the singular current $J_{B1}^{\left(0\right)}$ due to the additional factor
of $\ln z$. Instead, we may interpret this result as an operator mixing, requiring us to introduce an additional singular operator. At higher orders we expect
contributions of the form $1/z\times \ln^n z$ and a $d$-dimensional 
$z$-dependence of the form $z^{-1-n\epsilon}$. This prompts us to consider a tower of singular operators, defined in $d$ dimensions as
\begin{equation}
J_{B1}^{\left(n\right)}=\overline{\chi}_{c}\left(0\right)\gamma_{\mu}\left[(in_{-}\partial)^{1+n\epsilon}\mathcal{A^{\mu}}_{\perp\overline{hc}}\left[\left(\frac{1}{in_{-}\partial}\right)^{1+n\epsilon}\chi_{\overline{hc}}\right]\right]\left(0\right)\,.
\end{equation}
The off-shell regulated one-loop matrix element, including the sum of collinear, anti-hardcollinear and softcollinear loops is given by
\bea\label{eq:Jn1loop}
\lefteqn{ \langle  \bar q_{\overline{hc}}(p_1)g_{\overline{hc}}(p_2)| J_{B1}^{(n)}| \bar q_{ c}(p)\rangle_{\rm 1-loop} =}
\nn\\
&=& \frac{\alpha_s}{2\pi}\frac{1}{\epsilon^2}\,
\Bigg\{\T_1\cdot\T_0\left[\left(\frac{\mu^2zQ^2}{(-p_1^2)(-p^2)}\right)^{\!\epsilon}-\left(\frac{\mu^2}{-p_1^2}\right)^{\!\epsilon}-\left(\frac{\mu^2}{-p^2}\right)^{\!\epsilon}\,\right]\nn\\
  && +\T_2\cdot\T_0\left[\left(\frac{\mu^2\bar zQ^2}{(-p_2^2)(-p^2)}\right)^{\!\epsilon}-\left(\frac{\mu^2}{-p_2^2}\right)^{\!\epsilon}-\left(\frac{\mu^2}{-p^2}\right)^{\!\epsilon}\,\right]\nn\\
  && + \T_1\cdot\T_2\left(\frac{\mu^2}{-p_2^2}\right)^\epsilon\left[\frac{1}{z^\epsilon}-1\right]\Bigg\}\,\frac{1}{z^{1+n\epsilon}} \,g_s\bar v_{c}(p)\slashed{\epsilon}_{a\perp\overline{hc}}t^a v_{\overline{hc}}(p_1)
+{\cal O}\left(z^{0}\right)\,,
\eea
where we used the colour neutrality relation $\sum\T_i=0$ to rearrange terms and kept only the leading double poles. The dependence on the off-shell regulator cancels in the coefficient of the logarithmically enhanced
single-pole part when expanding in $\epsilon$, as expected. In addition, we find a term involving $\T_1\cdot\T_2$. From this point we could proceed to derive a $Z$-factor for the
set of singular currents. However, since all $J_{B1}^{(n)}$ coincide for $\epsilon\to 0$ the mixing of these currents cannot be determined unambigously in this way. Therefore, we
instead consider the product of the $d$-dimensional \emph{bare} Wilson coefficient and the bare 
currents, and use that the sum $\sum_n C_n J_n$ has no UV poles for any IR regulated matrix element. For the present case this leads to the condition
\be\label{eq:CnJn}
  \sum_n\Big\{ C_{B1,{\cal O}(\alpha_s)}^{(n)}\langle   J_{B1}^{(n)}\rangle_{\rm tree}  +C_{B1,{\rm tree}}^{(n)}\langle  J_{B1}^{(n)}\rangle_{\rm 1-loop}\Big\}_{1/\epsilon^2\ {\rm and}\ \ln(X)/\epsilon\ {\rm poles}}=0+{\cal O}\left(z^{0}\right)\,.
\ee
When matching to QCD, the first summand captures the hard one-loop contribution. The analysis at the beginning of Section~\ref{sec:derivation} implies that
the relevant regions contributing for $z\to 0$ have virtuality $Q^2$ or $zQ^2$. We therefore make the \emph{ansatz}
\be
  \sum_n C_{B1,{\cal O}(\alpha_s)}^{(n)}\langle   J_{B1}^{(n)}\rangle_{\rm tree} 
  = \frac{\alpha_s}{2\pi}\frac{1}{\epsilon^2} \sum_n \left[c_n\left(\frac{\mu^2}{Q^2}\right)^{\!\epsilon} 
  + d_n \left(\frac{\mu^2}{zQ^2}\right)^{\!\epsilon}\, \right]C_{B1,{\rm tree}}^{(n)}\langle   J_{B1}^{(n)}\rangle_{\rm tree}+{\cal O}\!\left(\frac{1}{\epsilon}\right).
\ee
Inserting this ansatz along with \eqref{eq:Jn1loop} into \eqref{eq:CnJn} yields $c_n=\T_2\cdot\T_0+\T_1\cdot\T_2=-C_A$ and $d_n=\T_1\cdot\T_0-\T_1\cdot\T_2=C_A-C_F$.

Since the building blocks in $J_{B1}^{(n)}$ are all evaluated at the same space-time position, the current operator itself cannot depend on $z$, and therefore the same applies to
the Wilson coefficients. Nevertheless, when evaluated in the matrix element, $\langle   J_{B1}^{(n)}\rangle_{\rm tree} \propto 1/z^{1+n\epsilon}$,
where $z$ is the momentum fraction of the external anti-hardcollinear quark. Therefore, the term involving the coefficient $d_n$ has to be interpreted as an operator
mixing $J_{B1}^{(n)}\to J_{B1}^{(n+1)}$, i.e.
\be
  C_{B1,{\cal O}(\alpha_s)}^{(n+1)} = \frac{\alpha_s}{2\pi}\frac{1}{\epsilon^2}\left(\frac{\mu^2}{Q^2}\right)^\epsilon \left[c_{n+1}C_{B1,{\rm tree}}^{(n+1)}+d_nC_{B1,tree}^{(n)}\right]+{\cal O}\!\left(\frac{1}{\epsilon}\right)\,,
\ee
implying for the ${\cal O}\left(\alpha_s\right)$ cusp part of the anomalous dimension matrix
\be
  \Gamma_{nm}^{B1,{\rm cusp}} = \frac{\alpha_s}{2\pi}\ln\frac{\mu^2}{Q^2} \times \left\{
  \begin{array}{ll}
   -C_A & \quad n=m\,,\\
   C_A-C_F & \quad n=m+1\,,\\
   0 & \quad {\rm else}\,.
  \end{array}\right. 
\ee
Solving the $d$-dimensional RG evolution (see 
Section~\ref{sec:T2qLPhard}) for the corresponding $Z$-factor,
\be
  \frac{d}{d\ln\mu^2}{\bf Z}=-{\bf Z}{\bf\Gamma}\,,
\ee
yields the bare Wilson coefficients $C_{B1}^{(n)}=Z_{nm}(Q)C_{B1,{\rm ren}}^{(m)}(Q)$ with 
\bea
  Z_{nm}(Q)=\exp\left[-C_A\frac{\alpha_s(\mu)}{2\pi}\frac{1}{\epsilon^2}\left(\frac{\mu^2}{Q^2}\right)^{\!\epsilon}\,\right]
  \sum_{j\geq 0}\frac{1}{j!}\left((C_A-C_F)\frac{\alpha_s}{2\pi}\frac{1}{\epsilon^2}\left(\frac{\mu^2}{Q^2}\right)^{\!\epsilon}\,
\right)^{\!j}\delta_{n,m+j}\,,\nn\\[-0.4cm]
\eea
where $\delta_{n,m+j}$ is the Kronecker symbol. At LL accuracy it 
is sufficient to evaluate the renormalized Wilson coefficients 
$C_{B1,{\rm ren}}^{(m)}(Q)$ at tree-level, such that all of them 
are zero, except for $m=0$, $C_{B1,{\rm ren}}^{(0)}(Q)=
C_{B1,{\rm tree}}^{(0)}=\kappa$. In the previous equation, 
we can therefore set $j=n$ and drop the sum over $j$. 

We are interested in the matrix element for $q\phi^*\to qg$, with leading poles arising from \emph{hard loops only}, which
is given by the hard matching coefficients multiplied with the tree-level SCET matrix element,
\bea
  \lefteqn{{\cal M}_{q\phi^*\to qg}\Big|_{\rm hard\ loops\ only} = \sum_n C_{B1}^{(n)}\langle  J_{B1}^{(n)} \rangle_{\rm tree} +{\cal O}(z^0)}\nn\\
  &=& C_{B1,{\rm tree}}^{(0)}\exp\left[-C_A\frac{\alpha_s}{2\pi}\frac{1}{\epsilon^2}\left(\frac{\mu^2}{Q^2}\right)^{\epsilon}+(C_A-C_F)\frac{\alpha_s}{2\pi}\frac{1}{\epsilon^2}\left(\frac{\mu^2}{z Q^2}\right)^{\epsilon}\right]\nn\\
 && \times\,\frac{1}{z}\, g_s\bar v_{ c}(p)\slashed{\epsilon}_{a\perp\overline{hc}}t^a v_{\overline{hc}}(p_1)+{\cal O}(z^0)\,.
\eea
Inserting this result into \eqref{eq:defPqg} precisely yields the exponentiation conjecture \eqref{eq:conjectureqghard}.


\section{Relation between DIS at large $x$ 
and event shapes in the two-jet limit}
\label{app:eventshapes}

In this Appendix we discuss the relation between NLP contributions 
to DIS for large $x$ and the thrust distribution in 
$e^+e^-\to\gamma^*(Q)\to$ jets \cite{Moult:2019uhz}. 
In particular, we consider the power expansion in the two-jet limit $\tau=1-T\to 0$, where $T$ is
the thrust event-shape variable, such that $\tau$ plays the role of $1-x$ (or $1/N$ in Mellin space)
in DIS.
The leading logarithmic corrections to the
differential cross section at NLP have the form~\cite{Moult:2016fqy}
\bea\label{eq:structureNLPLL}
  \frac{1}{\sigma_0}\frac{d\sigma}{d\tau}\Big|^{NLP,LL} &=& \sum_n \left(\frac{\alpha_s(Q)}{4\pi}\right)^n\, c_{LL}^{(n)}\,\ln^{2n-1}\tau\,.
\eea
The relevant regions are hard, (anti-)hardcollinear and soft, with virtualities $Q^2$, $\tau Q^2$ and $\tau^2Q^2$, respectively.
The leading poles can therefore be expanded in the form
\be\label{eq:structureNLPleadingpole}
  \frac{1}{\sigma_0}\frac{d\sigma}{d\tau}\Big|^{NLP,LL} = \sum_n \left(\frac{\alpha_s}{4\pi}\right)^n \frac{1}{\epsilon^{2n-1}}\sum_{j=1}^{2n} c_j^{(n)} \left(\frac{\mu^{2n}}{Q^{2n}\tau^j}\right)^\epsilon\,.
\ee
Hard loops contribute a factor $\alpha_s\times(\mu^2/Q^2)^\epsilon$, (anti-)hardcollinear loops $\alpha_s\times(\mu^2/Q^2\tau)^\epsilon$
and soft loops $\alpha_s\times(\mu^2/Q^2\tau^2)^\epsilon$. At NLP at least one (anti-)hardcollinear or soft loop is required, such that the expansion starts at $j=1$.
Compared to DIS, the virtualities are composed of only two independent scales instead of three, and therefore the coefficients $c_j^{(n)}$ depend only on a single index $j$ at each order in $\alpha_s$.
This implies that each coefficient can receive contributions from different combinations of regions, for example $c_2^{(2)}$ from one hard and one soft loop or two hardcollinear loops.
Pole cancellation yields $2n-1$ conditions
\be\label{eq:cnjconstraint}
  \left(\begin{array}{ccccc}
  1 & 1 & 1 & \dots & 1 \\
  1 & 2 & 3 & \dots & 2n \\
  1 & 4 & 9 & \dots & (2n)^2 \\
  \vdots\\
  1 & 2^{2n-2} & 3^{2n-2} & \dots & (2n)^{2n-2} \\
  \end{array}\right)\cdot \left(\begin{array}{c} c_1^{(n)}\\ c_2^{(n)}\\ \vdots \\ c_{2n}^{(n)}\end{array}\right)
  =\left(\begin{array}{c} 0\\ 0\\ 0\\ \vdots \\ 0\end{array}\right)\,,
\ee
that are all linearly independent. This means only one of the $2n$ coefficients $c_j^{(n)}$ is free. As noticed in~\cite{Moult:2016fqy}, the finite part
of \eqref{eq:structureNLPleadingpole} is completely determined by a single coefficient, specifically 
\be
  c^{(n)}_{LL} = -\frac{1}{(2n-1)!}\sum_{j=1}^{2n}  \,j^{2n-1}\,c_j^{(n)} =c_1^{(n)}\,,
\ee
where the last equality follows from solving \eqref{eq:cnjconstraint}. 
The only possible combination of regions contributing to $c_1^{(n)}$ are $n-1$ hard loops and one (anti-)hardcollinear loop. More precisely, the (anti-)hardcollinear loop arises from a phase-space integration
with a three-particle final state $\gamma^*\to q\bar q g$, with two of them being either both hardcollinear or both anti-hardcollinear.
We note that, for the thrust distribution, the hardcollinear and anti-hardcollinear directions have equal virtuality, and both refer to particles in the \emph{final} state.
Therefore, we are free to choose a convention for the light-cone basis such that $c_1^{(n)}$ receives contributions from $n-1$ hard loops and one \emph{anti}-hardcollinear loop.
This choice is made to make the analogy to DIS as close as possible, see below.

In the following we focus exclusively on those contributions to the NLP cross section for which no analog at LP exists (termed category II in~\cite{Moult:2016fqy}), in analogy to the off-diagonal DIS process. Category II requires either an anti-hardcollinear $q\bar q$ pair (IIc) or a soft $q$ or $\bar q$ (IIs), respectively. In the following it is understood that $c_j^{(n)}$ refers to category II only,
assuming that poles cancel separately in each category. Then $c_1^{(n)}$ receives only contributions from virtual hard loop corrections to $\gamma^*\to [q\bar q] g$, where the square bracket denotes
the anti-hardcollinear particles. Such contributions are given by
\be\label{eq:dsigdtau}
  \frac{1}{\sigma_0}\frac{d\sigma}{d\tau}\Big|_{\gamma^*\to [q\bar q] g} =   \int_0^1 dz\, \left(\frac{\mu^2}{s_{q\bar q}z\bar z}\right)^{\epsilon}{\cal P}_{q\bar q}(s_{q\bar q},z)\Big|_{s_{q\bar q}=Q^2\tau}+{\cal O}(\lambda^2) \,,
\ee
where $z$ and $\bar z$ are the collinear momentum fractions, $s_{q\bar q}$ is the virtuality of the $q\bar q$ pair, and
\be
  {\cal P}_{q\bar q}(s_{q\bar q},z)\equiv \frac{e^{\gamma_E\epsilon}Q^2}{16\pi^2\Gamma(1-\epsilon)}\frac{|{\cal M}_{\gamma^*\to [q\bar q] g}|^2}{|{\cal M}_0|^2}\,,
\ee
where $|{\cal M}_0|^2$ is the LO matrix element squared  
for $\gamma^*\to q\bar{q}$, and $|{\cal M}_{\gamma^*\to [q\bar q] g}|^2$ involves an arbitrary number of hard loop corrections.
This expression can be compared to (\ref{eq:TphiqVsz}) for DIS, which has a similar structure, except that here we consider a $1\to 3$ instead of a $2\to 2$ process,
and the anti-hardcollinear particles are a quark and an antiquark instead of a quark and a gluon. More importantly, DIS involves an additional scale related to the PDF, that is absent for
the thrust distribution. Nevertheless, as for DIS, the tree-level momentum distribution
\be
\label{eq:Pthrusttree}
  {\cal P}_{q\bar q}(s_{q\bar q},z)|_{tree} = \frac{\alpha_sC_F}{2\pi}\frac{e^{\gamma_E\epsilon}(1-\epsilon)}{\Gamma(1-\epsilon)}\left(\frac{\bar z}{z}+\frac{z}{\bar z}\right)\,,
\ee
leads to endpoint divergences in \eqref{eq:dsigdtau} for $z\to 0,1$. Using the conjecture from~\cite{Moult:2019uhz} for the all-order expression for ${\cal P}_{q\bar q}$ allows one to perform the $z$-integration
in $d$ dimensions, and, after expanding in $\alpha_s$, one can read off the coefficients $c_1^{(n)}$. Remarkably, the result coincides with (\ref{eq:cn1n}) multiplied by a factor
of two. Using $c^{(n)}_{LL} =c_1^{(n)}$ one directly obtains the LL contributions to the NLP thrust distribution from this result, which coincides with the ``soft quark Sudakov''
factor given in~\cite{Moult:2019uhz}.

\begin{figure}[t]
\vskip-0.3cm
\begin{center}
\includegraphics[width=0.30\textwidth]{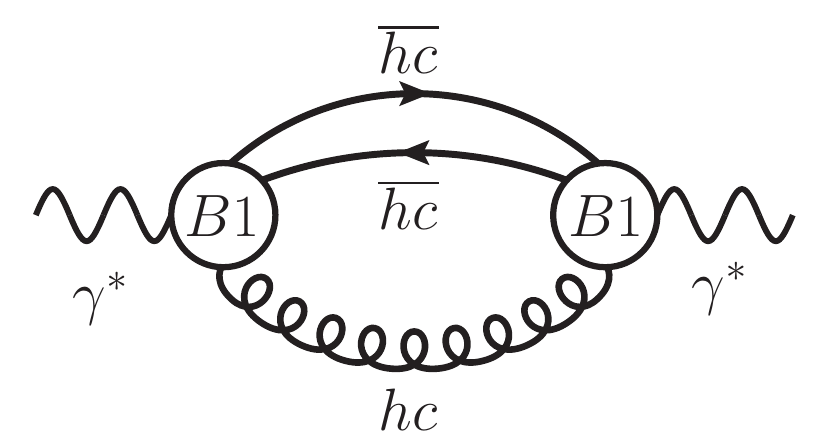}
\hspace*{1.5cm}
\includegraphics[width=0.30\textwidth]{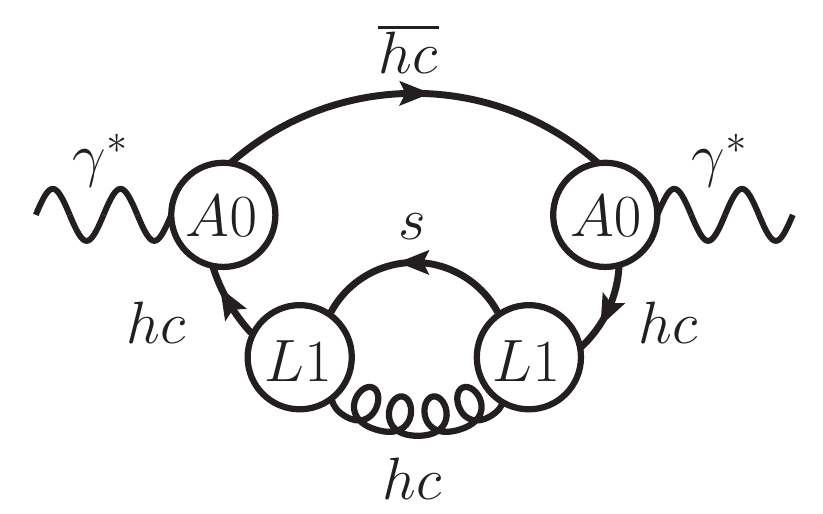}
\end{center} 
\vskip-0.3cm
\caption{SCET representation of the content of  
(\ref{eq:IIsIIc}) for the thrust distribution in $e^+e^-\to\gamma^*(Q)\to$ jets 
at NLP as $\tau\to 0$. Wilson lines are set to 1.
\label{fig:NLPSCETrepthrust}
}
\end{figure}

At this point one may wonder why, despite of the similarities, the LL resummed off-diagonal DGLAP kernel (\ref{eq:gamgqsolution}) obtained from the DIS process is considerably more complex than the
thrust distribution. To understand this difference, it is useful to separately consider contributions with an anti-hardcollinear $q\bar q$ pair (denoted by IIc) and those with 
a soft quark or antiquark (denoted by IIs).
The tentative SCET interpretation given
in~\cite{Moult:2019uhz} suggests that IIc is represented by diagrams involving a B1 current
operator with anti-hardcollinear $q\bar q$ building blocks, and IIs by diagrams with an insertion of a time-ordered product operator involving the LP current and ${\cal L}_{\xi q}^{(1)}$ (see Figure \ref{fig:NLPSCETrepthrust}).
This motivates the following ansatz for the (partial) factorization of hard, (anti-)hardcollinear and soft loop contributions to IIs and IIc,
\bea\label{eq:IIsIIc}
   \frac{1}{\sigma_0}\frac{d\sigma}{d\tau}\Big|^{NLP,LL}_{II} 
  &\equiv&  H^{LP}_{IIs}\cdot\left[J\times S\right]^{NLP}_{IIs} + \left[H\times J\right]^{NLP}_{IIc}\cdot S^{LP}_{IIc} \,,
\eea
with factorized hard and soft functions for IIs and IIc, respectively. They are governed by the usual LP cusp anomalous dimension, with leading poles given by
\bea
  H^{LP}_{IIs} &\equiv&  \exp\bigg[-\frac{\alpha_sC_F}{\pi}\frac{1}{\epsilon^2} \left(\frac{\mu^2}{Q^2}\right)^{\!\epsilon}\bigg]\,,\nn\\
  S^{LP}_{IIc} &\equiv&  \exp\bigg[-\frac{\alpha_sC_A}{\pi}\frac{1}{\epsilon^2} \left(\frac{\mu^2}{Q^2\tau^2}\right)^{\!\epsilon}\bigg]\,.
\eea
The non-trivial information resides in the combined jet and hard function for IIc, involving a convolution in momentum fractions of the B1 operator,
as well as in the combined soft and jet function for IIs, involving convolutions related to the spatial separation of the A0 current and the Lagrangian insertion. 
Following the discussion above, we expect these convolutions to feature endpoint divergences in four dimensions. 
The decomposition is analogous to (\ref{eq:Tpgifi}) in DIS, with $\left[H\times J\right]^{NLP}_{IIc}$  corresponding to the bare NLP partonic cross section $W_{\phi,q}^{NLP}$
and $\left[J\times S\right]^{NLP}_{IIs}$ to the bare NLP PDF evolution factor~$U^{NLP}_{gq}$.

Here we point out that, using the results from above allows one to bootstrap the resummed leading
poles of $\left[H\times J\right]_{IIc}$ and $\left[J\times S\right]_{IIs}$ in $d$ dimensions. To see this, we note that 
the consistency conditions \eqref{eq:cnjconstraint} determine all coefficients  $c_j^{(n)}$ for $1\leq j\leq 2n$ given the result for $c_1^{(n)}$. 
In addition, we use that the leading poles can be expanded in the form
\bea\label{eq:fdef}
  \left[H\times J\right]_{IIc} &=& \epsilon\,\sum_{n_c\geq 1}\left(-\frac{\alpha_s}{\pi}\frac{1}{\epsilon^2} \left(\frac{\mu^2}{Q^2\tau}\right)^\epsilon \right)^{n_c}\sum_{n_h\geq 0}
  \left(-\frac{\alpha_s}{\pi}\frac{1}{\epsilon^2} \left(\frac{\mu^2}{Q^2}\right)^\epsilon \right)^{n_h}c_{IIc}(n_c,n_h)\,,\nn\\
  \left[J\times S\right]_{IIs} &=& \epsilon\,\sum_{n_c\geq 0}\left(-\frac{\alpha_s}{\pi}\frac{1}{\epsilon^2} \left(\frac{\mu^2}{Q^2\tau}\right)^\epsilon \right)^{n_c}\sum_{n_s\geq 1}
  \left(-\frac{\alpha_s}{\pi}\frac{1}{\epsilon^2} \left(\frac{\mu^2}{Q^2\tau^2}\right)^\epsilon \right)^{n_s}c_{IIs}(n_c,n_s)\,,\nn\\[-0.4cm]
\eea
where $n_{s,c,h}$ denote the number of soft, (anti-)hardcollinear and hard loops. Note that $n_c>0$ for IIc and $n_s>0$ for IIs. Inserting this expansion into \eqref{eq:IIsIIc} 
and requiring that the sum of IIs and IIc contributions has to reproduce \eqref{eq:structureNLPleadingpole} with known coefficients $c_j^{(n)}$ allows one to
uniquely determine the coefficients $c_{IIc}(n_c,n_h)$ and $c_{IIs}(n_c,n_s)$. We find
\bea\label{eq:thrust_single}
   \left[H\times J\right]_{IIc}
  &=& \frac{C_F}{C_F-C_A}\frac{\epsilon\tau^{-\epsilon}}{\tau^{-\epsilon}-1} \nn\\
  && \Bigg\{ \exp\bigg[\frac{2\alpha_sC_A}{\pi}\frac{1}{\epsilon^2} \left(\frac{\mu^2}{Q^2\tau}\right)^\epsilon -\frac{\alpha_sC_A}{\pi}\frac{1}{\epsilon^2} \left(\frac{\mu^2}{Q^2}\right)^\epsilon\bigg]\nn\\
  && - \exp\bigg[ \frac{\alpha_s(C_F+C_A)}{\pi}\frac{1}{\epsilon^2} \left(\frac{\mu^2}{Q^2\tau}\right)^\epsilon -\frac{\alpha_sC_F}{\pi}\frac{1}{\epsilon^2} \left(\frac{\mu^2}{Q^2}\right)^\epsilon\bigg]\Bigg\}\,,\\
  \left[J\times S\right]_{IIs}
  &=& \frac{C_F}{C_F-C_A}\frac{\epsilon\tau^{-\epsilon}}{\tau^{-\epsilon}-1} \nn\\
  && \Bigg\{ -\exp\bigg[\frac{2\alpha_sC_F}{\pi}\frac{1}{\epsilon^2} \left(\frac{\mu^2}{Q^2\tau}\right)^\epsilon -\frac{\alpha_sC_F}{\pi}\frac{1}{\epsilon^2} \left(\frac{\mu^2}{Q^2\tau^2}\right)^\epsilon\bigg]\nn\\
  && + \exp\bigg[ \frac{\alpha_s(C_F+C_A)}{\pi}\frac{1}{\epsilon^2} \left(\frac{\mu^2}{Q^2\tau}\right)^\epsilon -\frac{\alpha_sC_A}{\pi}\frac{1}{\epsilon^2} \left(\frac{\mu^2}{Q^2\tau^2}\right)^\epsilon\bigg]\Bigg\}\,.
\label{eq:thrust_single2}\qquad
\eea
These expressions can be compared to  (\ref{eq:TphiqNLPsol}) for $W^{NLP,LP}_{\phi,q}$ and (\ref{eq:UphiqNLPsol}) for $U^{NLP,LP}_{gq}$, respectively. 
In particular, the last lines in each expression would lead to the appearance of Bernoulli functions when expanding the thrust distribution 
in $\epsilon$. Remarkably, however, these terms precisely cancel when adding the IIc and IIs pieces in \eqref{eq:IIsIIc}. The remaining terms combine to
exponential factors that are finite by themselves for $\epsilon\to 0$, giving
\bea
     \frac{1}{\sigma_0}\frac{d\sigma}{d\tau}\Big|^{NLP,leading\ poles}_{II}  
  &=& \frac{C_F}{C_F-C_A}\frac{\epsilon\tau^{-\epsilon}}{\tau^{-\epsilon}-1} \Bigg\{  \exp\bigg[-\frac{\alpha_sC_F}{\pi}\frac{1}{\epsilon^2} \left(\frac{\mu^2}{Q^2}\right)^{\!\epsilon}(1-\tau^{-\epsilon})^2\bigg]\nn\\
  && - \exp\bigg[-\frac{\alpha_sC_A}{\pi}\frac{1}{\epsilon^2} \left(\frac{\mu^2}{Q^2}\right)^{\!\epsilon}(1-\tau^{-\epsilon})^2\bigg]\,\Bigg\}\,.
\eea
This expression indeed has no poles in $1/\epsilon$, and approaches a finite limit for $\epsilon\to 0$, that precisely agrees with the
LL resummed ``soft quark Sudakov'' formula given in~\cite{Moult:2019uhz}.

While it is reassuring to recover the result for the LL resummed NLP thrust distribution given in~\cite{Moult:2019uhz}, the main purpose of this appendix is to point out the form of the two individual contributions \eqref{eq:thrust_single}, \eqref{eq:thrust_single2} and the formal analogy as well as difference to the DIS process.


\providecommand{\href}[2]{#2}\begingroup\raggedright\endgroup


\end{document}